\documentclass[11pt,a4paper]{article} 
\pdfoutput=1 
\usepackage{jheppub}
\usepackage{mciteplus}

\usepackage{graphicx}% Include figure files
\usepackage{dcolumn}% Align table columns on decimal point
\usepackage{mathrsfs}
\usepackage{slashed}

\def\be{\begin{equation}}
\def\ee{\end{equation}}
\def\beq{\begin{equation}}
\def\eeq{\end{equation}}
\def\bc{\begin{center}}
\def\ec{\end{center}}
\def\bea{\begin{eqnarray}}
\def\eea{\end{eqnarray}}

\newcommand{\eV}{\;\text{eV}}

\newcommand{\mbD}{\mathbf{D}}
\newcommand{\mbTh}{\mathbf{\Theta}}

\newcommand{\lhood}{\mathcal{L}}

\newcommand{\ie}{\emph{i.e.}}

\newcommand{\df}{{\rm d}}
\newcommand{\qu}[1]{``#1''}
\newcommand{\eref}[1]{Eq.~(\ref{#1})}

\def\T2K{{\sc T2K }}

\newcommand{\refcite}[1]{Ref.~\cite{#1}}
\newcommand{\figref}[1]{Fig.~\ref{#1}}
\newcommand{\tabref}[1]{Tab.~\ref{#1}}
\newcommand{\secref}[1]{Sec.~\ref{#1}}

\newcommand{\Dc}{D_{\text{c}}}
\newcommand{\Dsbl}{D_{\text{SBL}}}

\newcommand{\upd}{\text{upd}}

\newcommand{\meff}{m_{\text{\rm eff}}}
\newcommand{\ms}{m_{\text{\rm s}}}
\newcommand{\dNeff}{\Delta N_\text{\rm eff}}

\newcommand{\Bupd}{\mathcal{B}_{10}^\text{upd}}

\begin{document}
%
%%%
%%%%%%%%%%%%%%%%%%% Title Page%%%%%%%%%%%%%%%%%%%%%%%%%
%%%
%

\subheader{\footnotesize\sc Preprint number: YITP-SB-14-21, ICCUB-14-054, UB-ECM-PF-14-80}
\title{Statistical tests of sterile neutrinos using 
cosmology and short-baseline data}

\author[a]{Johannes Bergstr\"om}
\author[b c]{M.~C.~Gonzalez-Garcia}
\author[a]{V. Niro}
\author[d]{J. Salvado}

\affiliation[a]{Departament d'Estructura i Constituents de la Mat\`eria and 
Institut  de Ciencies del Cosmos,\\ Universitat de Barcelona, Diagonal 647,
  E-08028 Barcelona, Spain}

\affiliation[b]{Instituci\'o Catalana de Recerca i Estudis
  Avan\c{c}ats (ICREA),\\ 
  Departament d'Estructura i Constituents de la
  Mat\`eria and Institut de Ciencies del Cosmos,\\ Universitat de
  Barcelona, Diagonal 647, E-08028 Barcelona,
  Spain}

\affiliation[c]{C.N.~Yang Institute for Theoretical Physics, State 
University of New York at Stony Brook,\\ Stony Brook, NY 11794-3840, USA}

\affiliation[d]{Wisconsin IceCube Particle Astrophysics Center (WIPAC) 
and Department of Physics,\\ 
University of Wisconsin, Madison, WI 53706, USA\\}

\emailAdd{bergstrom@ecm.ub.edu}
\emailAdd{concha@insti.physics.sunysb.edu}
\emailAdd{niro@ecm.ub.edu}
\emailAdd{jordi.salvado@icecube.wisc.edu}

\abstract{
In this paper we revisit the question of the information which
cosmology provides on the scenarios with sterile neutrinos invoked to
describe the SBL anomalies using Bayesian statistical tests. 
We perform an analysis of the cosmological data 
in $\Lambda$CDM$+r+\nu_s$ cosmologies 
for different cosmological data combinations, and obtain the 
marginalized cosmological likelihood in terms of the two relevant
parameters, the sterile neutrino mass $m_s$ and its contribution
to the energy density of the early Universe $\dNeff$. 
We then present an analysis to quantify at which level a model
with one sterile neutrino is (dis)favoured with respect to a model with
only three active neutrinos, using results from both short-baseline
experiments and cosmology. We study the dependence of the results
on the cosmological data considered, in particular on the inclusion
of the recent BICEP2 results and the SZ cluster data from the Planck
mission. We find that only when the cluster data is included the 
model with one extra sterile neutrino can become more favoured that 
the model with only the three active ones provided the sterile neutrino
contribution to radiation density is suppressed with respect 
to the fully  thermalized scenario.  
We have also quantified the level of (in)compatibility between the
sterile neutrino masses implied by the cosmological and SBL results.  
}

\maketitle

%---------------------------------------------------------------------
\section{Introduction}
%---------------------------------------------------------------------

It is now an established fact that neutrinos are massive and leptonic
flavors are not symmetries of Nature~\cite{Pontecorvo:1967fh,
Gribov:1968kq}.  In the last decade this picture has become fully
established thanks to the upcoming of a set of precise experiments. In
particular, the results obtained with solar and atmospheric neutrinos
have been confirmed in experiments using terrestrial
beams~\cite{GonzalezGarcia:2007ib}.  The minimum joint description of
all these data requires mixing among all the three known neutrinos
($\nu_e$, $\nu_\mu$, $\nu_\tau$), which can be expressed as quantum
superposition of three massive states $\nu_i$ ($i=1,2,3$) with masses
$m_i$ leading to the observed oscillation signals with $\Delta
m^2_{21}= (7.5^{+0.19}_{-0.17})\times 10^{-5}$ eV$^2$ and $|\Delta
m^2_{31}|=(2.458\pm 0.002)\times 10^{-3}$ eV$^2$ and non-zero values
of the three mixing angles \cite{GonzalezGarcia:2012sz}.

In addition to these well-established results, there remains a set of anomalies
in neutrino data at relatively short-baselines (SBL) (see \refcite
{Abazajian:2012ys} for a review) including 
the LSND~\cite{LSND:2001ty} and MiniBooNE
~\cite{MiniBooNE:2010wv,Aguilar-Arevalo:2013pmq} observed 
$\nu_\mu\rightarrow \nu_e$ transitions, and  the $\bar\nu_e$ disappearance
at reactor~\cite{Mention:2011rk,Mueller:2011nm,Huber:2011wv} and
Gallium~\cite{Giunti:2010zu,Giunti:2012tn,Acero:2007su}
experiments. If interpreted in terms of oscillations, each of these anomalies 
points out towards a $\Delta m^2\sim {\cal O}({\rm eV}^2)$ 
~\cite{Kopp:2011qd,Giunti:2011gz,Giunti:2011hn,Giunti:2011cp,Donini:2012tt,
Conrad:2012qt,Giunti:2013aea,Karagiorgi:2012kw,Kopp:2013vaa} 
and consequently cannot be explained within the context of the 3$\nu$
mixing described above. They require instead the addition of one or
more neutrino states which must be {\sl sterile}, {\it i.e.} elusive
to Standard Model interactions, to account for the constraint of the
invisible $Z$ width which limits the number of light weak-interacting
neutrinos to be $2.984 \pm 0.008$~\cite{Nakamura:2010zzi}. 

Several combined analyses have been performed to globally account for
these anomalies in addition to all other oscillation results in the
context of models with one or two additional sterile neutrinos
\cite{Giunti:2013aea,Karagiorgi:2012kw,Conrad:2012qt,Kopp:2013vaa} with somehow
different conclusions in what respects to the possibility of a
successful joint description of all the data.  
Generically these global fits reveal a tension between disappearance and 
appearance results. 
But while Refs.~\cite{Giunti:2013aea,Karagiorgi:2012kw} 
seem to find a possible compromise solution for 3+1 mass schemes, 
the analysis in 
\refcite{Kopp:2013vaa} concludes a significantly lower level of 
compatibility. In particular, \refcite{Giunti:2013aea} concludes
that a joint solution is found with 
\begin{eqnarray} 
0.82\leq \Delta m^2_{41} \leq  2.19 \;{\rm eV}^2\;\;  {\rm at}\rm\;\; 3\sigma \\
\nonumber
|U_{e4}|^2\sim 0.03\, ,  \;\;\; |U_{\mu 4}|^2\sim 0.012\;.
\label{eq:parsbl} 
\end{eqnarray} 

Alternative information on the presence of light sterile neutrinos is
provided by Cosmology as they contribute as extra radiation to the
energy density of the early Universe which can be expressed as 
\be
\rho_R = \left[ 1 + \frac{7}{8} \left( \frac{4}{11} \right)^{4/3}
N_{\rm eff} \right] \rho_\gamma \,, 
\ee 
where $\rho_\gamma$ is the
photon energy density and the value of $N_{\rm eff}$ in the Standard
Model (SM) is equal to $N_{\rm eff}^{\rm SM}=3.046$~\cite{Mangano:2005cc}.
The presence of extra radiation is then usually quantified in terms of
the  parameter $\Delta N_{\rm eff} \equiv N_{\rm eff} - N^{\rm SM}_{\rm eff}$.

Sterile neutrinos contribute to $\rho_R$ ({\it i.e.} to $\Delta N_{\rm
eff}$) in a quantity which, in the absence of other forms of new
physics, is a function of their mass and their mixing with the active
neutrinos which determines to what degree they are in thermal
equilibrium with those.  In particular, in $3+N_s$ scenarios with the
values hinted by the SBL results (of the order of those in
\eref{eq:parsbl}) the sterile neutrinos
are fully thermalized (FT) and  each one contributes to $\rho_R$  as much as  
an active one (see for example \cite{Saviano:2013ktj,Mirizzi:2013kva}), so
\begin{equation}
\Delta N^{3+N_s,FT}_{\rm eff}=N_s \:. 
\label{eq:FT}
\end{equation}
 
Analyses of cosmological data have hinted for the presence of extra radiation,
beyond the standard three active neutrinos since several years (see
for example \refcite{GonzalezGarcia:2010un} and references therein) 
and several authors have
invoked the presence of eV-scale sterile neutrino as a plausible
source of the extra radiation ~\cite{Acero:2008rh,Hamann:2010bk,Giusarma:2011zq,Archidiacono:2011gq,Archidiacono:2012ri}\footnote{Alternative scenarios without sterile neutrinos have been proposed as well  (see for example
Refs.~\cite{GonzalezGarcia:2012yq,DiBari:2013dna,Hasenkamp:2012ii,Graf:2013xpe}
and references therein)}.
In the last four years the statistical significance of this extra radiation 
(or its upper bound) as well as the overall constraint on the neutrino mass
scale has been changing  as data from the Planck
satellite~\cite{Ade:2013lta}, the Atacama Cosmology Telescope (ACT)~
\cite{Das:2013zf}, the South Pole Telescope (SPT)~
\cite{Keisler:2011aw,Reichardt:2011yv}, and most recently of the
BICEP2~\cite{Ade:2014xna} experiment, has become available 
~\cite{Verde:2013cqa,Archidiacono:2014apa, Dvorkin:2014lea,Zhang:2014dxk, Zhang:2014nta, Wu:2014qxa, Leistedt:2014sia,Giusarma:2014zza}.

In particular, the recent measurement of B-mode signals from the
BICEP2~\cite{Ade:2014xna} collaboration excludes a zero
scalar-to-tensor ratio at $7\sigma$ and report a value of 
$r = 0.20^{+0.07}_{-0.05}$.
This high value of $r$ is compatible with previous results from the
Planck data~\cite{Ade:2013lta} if a running of the scalar
spectral index $d n_s/d \ln k$ is considered well beyond the characteristic
value of $10^{-4}$  of slow-roll inflation models. Alternatively 
the tension might be eased by the presence of sterile neutrinos
\cite{Archidiacono:2014apa, Dvorkin:2014lea,Zhang:2014dxk,Zhang:2014nta,Wu:2014qxa} without invoking such a large running of the scalar spectral index. 

Since cosmological data have the potential to test regions of
parameter space of the sterile neutrino scenarios invoked to account for
the  SBL anomalies, the question of to what degree they support them has
received an increasing attention in the literature~\cite{Hamann:2011ge,Archidiacono:2012ri,Kristiansen:2013mza,Archidiacono:2013xxa,Gariazzo:2013gua}.
The generic conclusion is that, in order to accommodate the cosmological
observations within the $3+N_s$ scenarios motivated by SBL results, some
new form of physics is required to suppress the contribution of 
the sterile neutrinos to the radiation component of the energy
density at the CMB epoch with respect to the FT expectation \eref{eq:FT}.
Among others extended scenarios with   
a time varying dark energy component~\cite{Giusarma:2011zq}, entropy
production after neutrino decoupling~\cite{Ho:2012br}, 
very low reheating temperature~\cite{Gelmini:2004ah}, large lepton
asymmetry~\cite{Foot:1995bm,Chu:2006ua,Saviano:2013ktj}, and non-standard
neutrino interactions ~\cite{Bento:2001xi,Dasgupta:2013zpn,Hannestad:2013ana},
have been considered.
All these mechanisms have the effect of diluting the sterile neutrino 
abundance or suppressing the production in the early Universe.

In this paper we revisit the question of the information which 
cosmology provides on the sterile scenarios introduced to explain 
the SBL anomalies using precise Bayesian statistical tests which 
we briefly describe in \secref{sec:statistical}. Section \ref{sec:cosmo}
contains the results of our cosmological analysis in 
$\Lambda$CDM$+r+\nu_s$ cosmologies for three representative sets of cosmological 
data. With these results at hand we answer the question of 
{\sl how much cosmological data
favour or disfavour the scenario with sterile neutrino masses
invoked by SBL anomalies, with respect to a model without sterile neutrinos?}
in \secref{sec:sterile},  and we do so in terms of the departure from the 
fully thermalized expectation, \eref{eq:FT}. We also discuss the 
consistency of sterile parameter constrains implied by cosmology and
SBL.  In \secref{sec:conclusions} we summarize our conclusions.

%---------------------------------------------------------------------
\section{The statistical framework}\label{sec:statistical}
%---------------------------------------------------------------------
\subsection{Model comparison}
Bayesian inference is a rigorous framework for inferring which, out of
set of models or hypotheses $H_i$, are favoured by a data set
$\mathbf{D}$.  Bayes' theorem is used to calculate the probabilities
of each of the hypotheses after considering the data, the
\emph{posterior probabilities},
\begin{equation}\label{eq:Bayes_model} 
\Pr(H_i|\mathbf{D}) = \frac{\Pr(\mathbf{D}|H_i)\Pr(H_i)}{\Pr(\mathbf{D})}.
\end{equation} 
Here $\Pr(\mathbf{D}|H_i)$ is the probability of the data, assuming
the model $H_i$ to be true, while $\Pr(H_i)$ is the \emph{prior
probability} of $H_i$, which is how plausible $H_i$ is before
considering the data. Considering especially the case of a discrete
set of models, one can compare two of them by calculating the ratio of
posterior probabilities, the \emph{posterior odds}, as
\begin{equation}\label{eq:post_ratio} 
\mathcal{O}_{ij}=\frac{ \Pr(H_i|\mathbf{D})}{\Pr(H_j|\mathbf{D})} =
\frac{\Pr(\mathbf{D}|H_i)}{\Pr(\mathbf{D}|H_j)} 
\frac{\Pr(H_i)}{\Pr(H_j)}= B_{ij} \frac{\Pr(H_i)}{\Pr(H_j)}.
\end{equation}
In words, the posterior odds is given by the \emph{prior odds} $
\Pr(H_i)/ \Pr(H_j)$ multiplied by the \emph{Bayes factor} $B_{ij} =
\Pr(\mathbf{D}|H_i)/{\Pr(\mathbf{D}|H_j)}$, which quantifies how much
better $H_i$ describes that data than $H_j$.  The prior odds
quantifies how much more plausible one model is than the other a
priory, \ie, without considering the data. If there is no reason to
favour one of the models over the other, the prior odds equals unity,
in which case the posterior odds equals the Bayes factor.

For a model  $H$, containing the continuous free parameters $\mbTh$, 
$\Pr(\mathbf{D}|H)$ also called \emph{evidence} of the model is given by
\bea
\Pr(\mathbf{D}|H) 
&=&  \notag  \\ 
\int \Pr(\mathbf{D},\mbTh|H)\df^N\mathbf{\Theta} &=&  
\int \Pr(\mathbf{D}|\mbTh, H) \Pr(\mathbf{\Theta}|H)\df^N\mathbf{\Theta} 
\notag \\
&=& \int{\mathcal{L}(\mathbf{\Theta})\pi(\mathbf{\Theta})}\df^N\mathbf{\Theta}.
\label{eq:Z}
\eea
Here, the \emph{likelihood function} $\Pr(\mathbf{D}|\mathbf{\Theta},
H)$ is the probability (density) of the data as a function of the
assumed free parameters, which we often denote by
$\mathcal{L}(\mathbf{\Theta})$ for simplicity.  The quantity
$\Pr(\mathbf{\Theta}|H)$ is the correctly normalized prior probability
(density) of the parameters and is often denoted by
$\pi(\mathbf{\Theta})$.  The assignment of priors is often far from
trivial, but an important part of a Bayesian analysis. 

From \eref{eq:Z}, we note that the evidence is the average of the
likelihood over the prior, and hence this method automatically
implements a form of \emph{Occam's razor}, since usually a theory with
a smaller parameter space will have a larger evidence than a more
complicated one, unless the latter can fit the data substantially
better. 

Bayes factors, or rather posterior odds, are usually interpreted or
\qu{translated} into ordinary language using the so-called
\emph{Jeffreys scale}, given in \tabref{tab:Jeffreys}, where
\qu{$\log$} is the natural logarithm.  This has been used in
applications such as
Refs.~\cite{Trotta:2008qt,Hobson:2010book,Feroz:2008wr} (and
Refs.~\cite{Bergstrom:2012yi,Bergstrom:2012nx} in neutrino physics),
although slightly more aggressive scales have been used previously
\cite{Jeffreys:1961,Kass:1995}.
\begin{table}
\begin{center}
\begin{tabular}{c|c|c|c}
\hline
$| \log(\text{odds}) |$ & odds & $\Pr(H | \mathbf{D})$ & Interpretation \\ 
\hline\hline
$<1.0$ & $\lesssim 3:1$ & $\lesssim 0.75$ & Inconclusive \\
$1.0$ & $\simeq 3:1$ &  $\simeq 0.75$ & Weak evidence \\
$2.5$ & $\simeq 12:1$ & $\simeq 0.92$ & Moderate evidence \\
$5.0$ & $\simeq 150:1$ & $ \simeq 0.993$ & Strong evidence \\ \hline
\end{tabular}
\end{center}
\caption{Jeffrey's scale often used for the interpretation of model
odds. The posterior model probabilities for the preferred model $H$
are calculated by assuming only two competing hypotheses.}
\label{tab:Jeffreys}
\end{table}

\subsection{Parameters of interest and the marginal likelihood}
\label{sec:marginalL}

In Bayesian statistics if one assumes a particular parametrized model to 
be correct, the complete inference of the parameters of that model is 
given by the posterior distribution through Bayes' theorem
\begin{equation}\label{eq:Bayes_params} 
\Pr( \mathbf{ \Theta} | \mathbf{D},H) = \frac{\Pr(\mathbf{D}
|\mathbf{\Theta},H)\Pr(\mathbf{\Theta}|H)}
{\Pr(\mathbf{D}|H)}  = \frac{\lhood(\mbTh)\pi(\mbTh)}{\Pr(\mathbf{D}|H)}.
\end{equation}
We see that the evidence here appears as a normalization constant in
the denominator.  Since the evidence does not depend on the values of
the parameters $\mbTh$, it is usually ignored in parameter estimation
problems and the parameter inference is obtained using the
unnormalized posterior. 

For the case in which we have a only a subset of parameters of interest,
$\lambda$,  so $\mbTh=(\lambda,\eta)$, where 
$\eta$ denotes the \emph{nuisance
parameters},
\be 
P(\lambda,\eta) = \Pr(\lambda,\eta|\mbD, H) \propto
\lhood(\lambda,\eta) \pi(\lambda,\eta)\,, 
\ee 
and inference on $\lambda$ is  obtained by marginalizing over the nuisance
parameters in the usual way 
\be 
P(\lambda) = 
\int P(\lambda,\eta) \df\eta\,, \label{eq:margpost} 
\ee 
with no need to ever consider a likelihood $\lhood(\lambda)$ 
depending only on the parameter of
interest.  This is typically unproblematic in the case where the data
is sufficiently informative to eliminate all practical prior
dependence. Often, however, this is not the case, and there can be
large dependence on the prior chosen.  
In this case one can consider as a partial step
the marginal likelihood function
\be \lhood(\lambda) = \int \lhood(\lambda,\eta)
\pi(\eta) \df \eta \,, 
\ee such that 
$P(\lambda) \propto \lhood(\lambda) \pi(\lambda).$

This likelihood function then encodes the information on $\lambda$
contained in the data (under $H$ and after taking into account the
uncertainty on the nuisance parameters), without needing to specify a
prior $\pi(\lambda)$.  Note that, since the marginal likelihood is not
a probability density, it is not normalized to unity, and is not
sufficient to perform the full inference. Also, it is generally
different from the profile likelihood, so regions defined by $ -2
\log(\lhood(\lambda)/\lhood^{\rm max})<C$ will not, in general, be the same as
those using the profile likelihood, although arguments justifying
defining regions in this way for profile likelihoods typically also apply
to marginal likelihoods. As shown in \refcite{Planck:2013nga}, the
profile and marginal likelihoods are indeed similar for the
cosmological data and models considered there.  In any case, the
marginal likelihood can still be useful in scientific reporting as a
rough guide to what information the data contains, for example, by
considering regions defined by $ -2 \log(\lhood(\lambda)/\lhood^{\rm
max})<C$ \cite{berger1999}.

Furthermore, if two data sets do not share any common nuisance
parameters, the two marginal likelihoods can simply be multiplied to
obtain the total marginal likelihood.  Notice, however, that the marginal
likelihood still depends on the priors on the nuisance parameters. If the
nuisance parameters are well-constrained this dependence will be
small, but in cosmology this is necessarily not always the case.

%---------------------------------------------------------------------
\section{Cosmological analysis}
\label{sec:cosmo}
%---------------------------------------------------------------------

\subsection{Data sets} 
\label{subsec:data}
In our cosmological analysis, we use data on Cosmic
Microwave Background, large scale structure (LSS)   
baryon acoustic oscillations (BAO) measurements, 
Hubble constant $H_0$, and galaxy cluster counts. 
In particular, we define the following data combinations:
\begin{itemize}
\item CMB: It includes the current Planck
data~\cite{Ade:2013lta} of the temperature anisotropy
up to $l=2479$, the high multipole values~(highL), coming from
ACT~\cite{Das:2013zf} and SPT
data~\cite{Keisler:2011aw,Reichardt:2011yv},
that covers respectively the $500 < l < 3500$ and $600 < l < 3000$ range,
and the EE and TE polarization data from WMPA9~\cite{Bennett:2012zja} (WP).
It also includes the CMB lensing potential (lensing)
reconstructed by the Planck collaboration~\cite{Ade:2013lta}
through the measurement of the four-point function. 
\item BAO: It includes the Data Release~11 (DR11) sample of the recent 
measurements by the BOSS
collaboration~\cite{Anderson:2013zyy}. The DR11 sample is the largest
region of the Universe ever surveyed,
covering roughly 8500 square degrees, with a redshift range $0.2<z<0.7$.
The measure of the sound horizon at the drag epoch has been evaluated
at redshift $z$=0.32 and at $z$=0.57, finding values in agreement with
previous BAO measurements \footnote{We have verified that the inclusion of the
previous determinations of the
BAO from the Sloan Digital Sky Survey (SDSS) Data Release 7 (DR7)
~\cite{Percival:2009xn,Padmanabhan:2012hf}, 
the 6dF Galaxy Survey (6dFGS)~\cite{Beutler:2011hx}, and  
WiggleZ measurements \refcite{Blake:2011en} does not affect our results
and therefore for the sake of simplicity we have not included them
in the analysis.}.
\item BICEP2: the 9 channels of the CMB BB polarization spectrum 
recently released by the BICEP2 experiment~\cite{Ade:2014xna}.
\item HST: the data from the Hubble Space
Telescope~\cite{Riess:2011yx} on $H_0$, obtained 
through the distance measurements of the Cepheids:
\be
H_0 = (73.8 \pm 2.4)~\text{km s}^{-1}\text{Mpc}^{-1}\,.
\ee
\item PlaSZ: The counts of rich cluster of galaxies from the sample of
Planck thermal Sunyaev-Zel'Dovich catalogue \cite{Ade:2013lmv}. It constrains
the combination $\sigma_8(\Omega_m/0.27)^{0.3} = 0.782 \pm 0.010$
\footnote{For simplicity we do not include 
the cosmic shear data of weak lensing from the Canada-French-Hawaii
Telescope Lensing Survey (CFHTLenS) which constraints 
$\sigma_8(\Omega_m/0.27)^{0.46} = 0.774 \pm 0.040$, 
~\cite{Benjamin:2012qp, Heymans:2013fya,Kilbinger:2012qz}. Although
the combinations constrained by PlaSZ and CFHTLenS 
are not the same, given the much better precision of the PlaSZ data, 
the impact of including CFHTLenS in our analysis is very small.}  
\end{itemize}

In order to test the dependence of our results on the inclusion of
the recent BICEP2 data and on the tension  with local HST and cluster 
PlaSZ results we perform the analysis with three different combinations of 
the data sets above that we label: 
\begin{itemize}
\item {\it DATA SET 1:} CMB+BAO, where, as described above,  
CMB=Planck+WP+highL+lensing data,  and BAO=DR11. 
\item {\it DATA SET 2:} CMB+BAO+BICEP2. We add to the previous data set, 
the results from BICEP2.  
\item {\it DATA SET 3:} CMB+BAO+BICEP2+HST+PlaSZ. 
We add to the previous data set the results from HST and Planck SZ counts 
of galaxy clusters. 
\end{itemize}

\subsection{Cosmological model}
We consider in our analysis a $\Lambda$CDM cosmology extended 
with a free  scalar-to-tensor ratio, and three active plus one sterile
neutrino species with a hierarchical neutrino spectra of the 3+1 type which
we denote as $\Lambda$CDM$+r+\nu_s$.
In this case the three active neutrinos have masses $m_{i=1,3}\lesssim
\sqrt{|\Delta m^2_{31}|}$ while the forth sterile neutrino 
has a mass $m_4\equiv m_s$. 

As mentioned in the introduction, in the absence of other form of 
new physics the contribution of the sterile neutrino to the energy density 
is completely determined by its mass and its mixing with the active neutrinos.
However in extended scenarios this may not be the case. So generically 
we will consider that in the 3+1 scenario, irrespective of $m_s$
and mixings,  the sterile neutrino contributes  to $\rho_R$ as 
\begin{equation}
\Delta N_{\rm eff}\equiv F_{NT} \, \Delta N^{3+1,FT}_{eff}=F_{NT}  
\label{eq:FNT}
\end{equation}      
where $F_{NT}$ is an arbitrary quantity which quantifies the departure
from the fully-thermalized active-sterile neutrino scenario and which 
we will consider to be independent of $T$ in the relevant range
of $T$ in the analysis. So in what follows we will label as 
$\Delta N_{\rm eff}\equiv F_{NT}$ this parameter.   

The effect of $m_s$ is included in the analysis via the effective parameter
\be 
m_{\rm eff} = (94.1~\text{eV})~\Omega_s h^2\,, 
\ee 
with being $h$ the reduced Hubble constant and $\Omega_s \equiv \rho_s/\rho_c$,
with $\rho_s$ the sterile neutrino energy density and $\rho_c$ the
current critical density. This effective mass is not equal to the physical 
mass $m_s$ in general, and their relation depends on the assumed phase-space 
distribution of the sterile neutrinos. For thermally  distributed 
sterile neutrinos characterized by a temperature $T_s$ (in general
different from the temperature of the active neutrinos  $T_\nu$) 
\be
m_{\rm eff}  = (\Delta N_{\rm eff})^{3/4}m_s \, , 
\label{eq:msther}
\ee
while if they are produced by non-resonant oscillations 
(the so-called Dodelson-Widrow scenario) (DW)~\cite{Dodelson:1993je} the
resulting  phase-space distribution of the sterile neutrinos is equal
to that of the active neutrinos up a constant factor. In this case  
\be 
m_{\rm eff}  = \Delta N_{\rm eff} m_s \,.
\ee 

Altogether our analysis contain nine free parameters
\be 
\{ \omega_b, \omega_c, \Theta_s, \tau, \log[10^{10} A_s], n_s,r,  
m_{\rm eff}, N_{\rm eff} \} \,, 
\label{eq:param}
\ee 
where $\omega_{\rm b}
\equiv \Omega_{\rm b} h^2$, $\omega_{\rm c} \equiv \Omega_{\rm c} h^2$
being the physical baryon and cold dark matter energy densities,
$\Theta_{\rm s}$ the ratio between the sound horizon and the angular
diameter distance at decoupling, $\tau$ is the reionization optical
depth, $A_{\rm s}$ the amplitude of primordial spectrum, $n_{\rm s}$
the scalar spectral index, and $r$ the scalar-to-tensor ratio. To generate
the marginalized likelihoods we use the \textsc{CosmoMC} 
package~\cite{Lewis:2002ah}, implemented with  the Boltzmann CAMB 
code~\cite{Lewis:1999bs}. 

The parameters in (\ref{eq:param}) are assigned uniform priors with limits 
as given in \tabref{tab:priors}. 
Since we are interested in the constraints on sterile neutrino
parameters, we follow \secref{sec:marginalL}, and aim to evaluate the
marginal likelihoods of ($m_{\rm eff}, N_{\rm eff}$), which will then
be used for the tests presented in \secref{sec:sterile}. Thus in this analysis 
we consider $\{ \omega_{\rm b}, \omega_{\rm c}, \Theta_{\rm s},
\tau, \log[10^{10} A_{\rm s}], n_{\rm s},r \}$ as
the {\sl cosmological nuisance parameters}
\footnote{We do not include the lensing amplitude $A_{\rm L}$ as a
free nuisance parameter, even when adding the local measurements on
$\sigma_8$ and $\Omega_m$, for which a significant deviation from the
standard value of unity is preferred. We have checked that
adding $A_{\rm L}$ only slightly shifts the preferred region for
$m_{\rm eff} $ to higher values.}. 
Note that we have employed uniform priors on $ m_{\rm eff}$ and
$N_{\rm eff}$ in the numerical analysis but that the marginal
likelihoods to be presented below do not depend on the priors, since 
these are \qu{factorized out}. Furthermore, since the nuisance parameters
are rather well-constrained the precise vales of the limits in
\tabref{tab:priors} do not not affect any results, and also employing
different shapes on these parameters would have a small impact.  The
exception is $r$ which is not so well-constrained (especially for
certain data sets) and for which the physical lower limit is
important.
\begin{table}
\begin{center}
\begin{tabular}{c|c}
\hline
Parameter & Prior \\ 
\hline\hline
$\Omega_b h^2$ & $0.005 \rightarrow 0.1$ \\
$\Omega_c h^2$ & $0.001 \rightarrow 0.99$ \\
$\Theta_s$ & $0.5 \rightarrow 10$ \\
$\tau$ & $0.01 \rightarrow 0.8$ \\
$\ln[10^{10} A_s]$ & $2.7 \rightarrow 4$ \\ 
$n_s$ & $0.9 \rightarrow 1.1$ \\ 
$r$ & $0 \rightarrow 1$ \\ \hline
$m_{\rm eff}$ & $0 \rightarrow 3$ \\  
$N_{\rm eff}$ & $3.046 \rightarrow 6$ \\  
\hline
\end{tabular}
\end{center}
\caption{Uniform priors for the cosmological parameters considered in
the analysis.  The active neutrinos have been fixed to one
massive with a mass of 0.06 and two massless. In addition 
following \refcite{Ade:2013lta} an upper
constraint on $m_s$ defined in \eref{eq:msther} of 10 eV is imposed
which roughly defines the
region where the sterile neutrinos are distinct from cold
or warm dark matter.}
\label{tab:priors}
\end{table}

\subsection{Marginal cosmological likelihoods} 
\begin{figure}
\begin{tabular}{cc}
\multicolumn{2}{c}{\large CMB+BAO} \\& 
Thermal $\nu_s$ \\
\includegraphics[width=0.5\textwidth,clip=true]{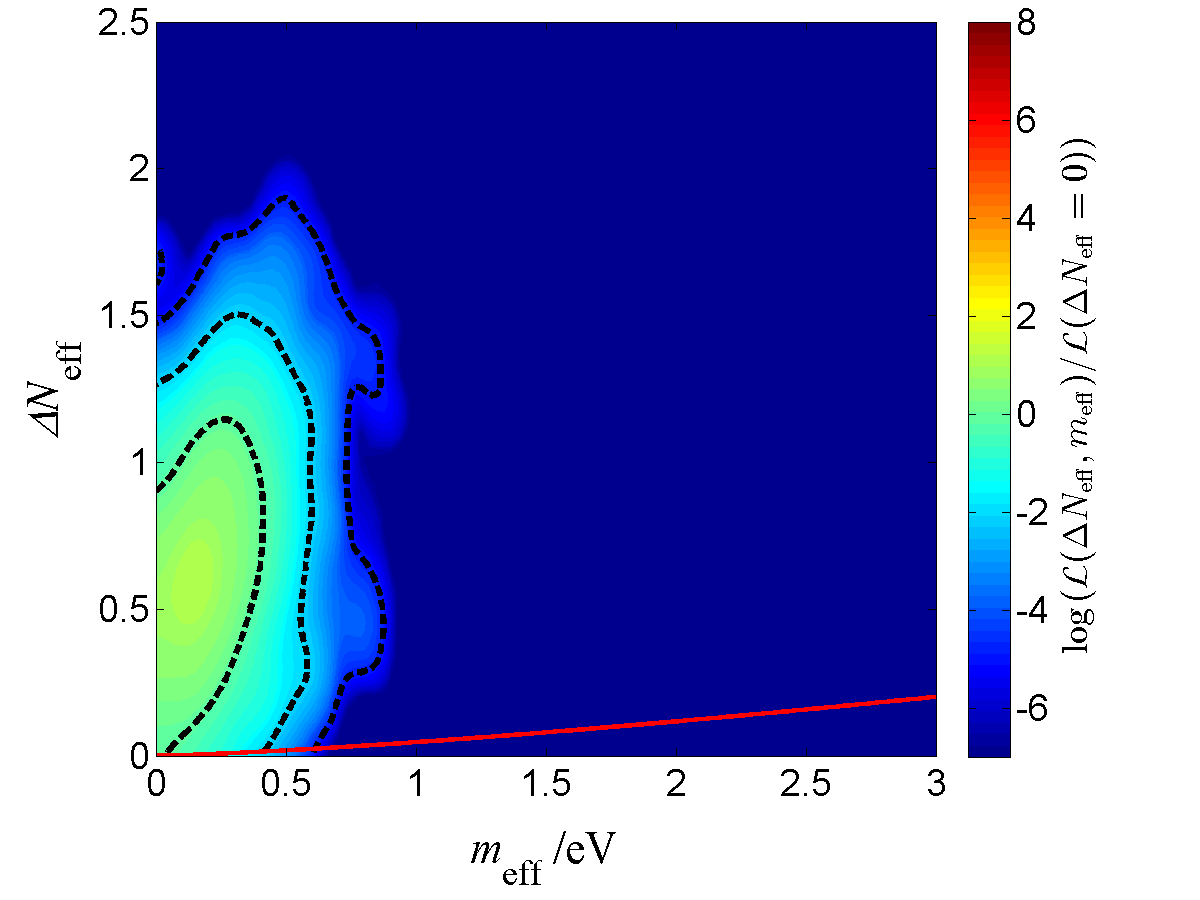} & 
\includegraphics[width=0.5\textwidth,clip=true]{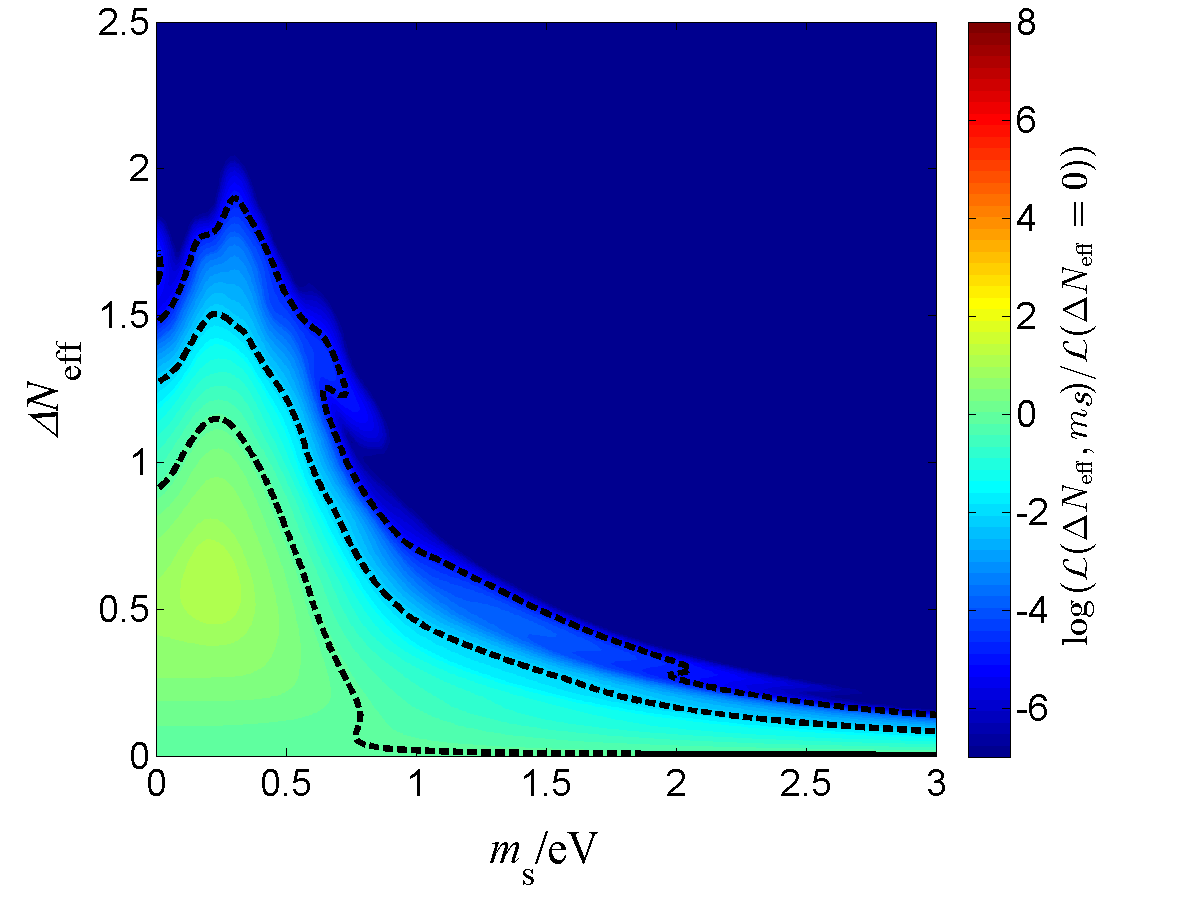} \\
Dodelson-Widrow scenario\\
\includegraphics[width=0.5\textwidth,clip=true]{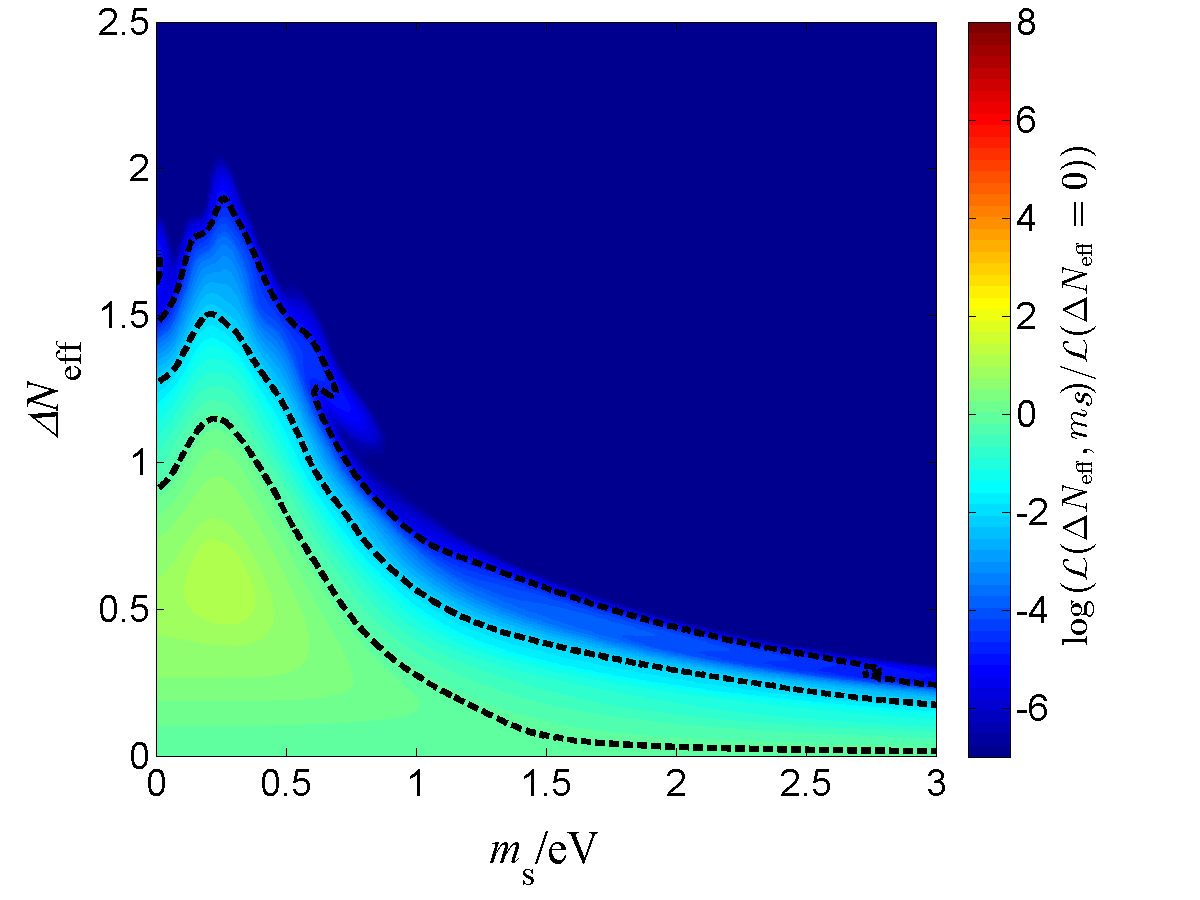}
\end{tabular}
\caption{Marginal likelihood of $(m_{\rm eff},\Delta N_{\rm eff})$
(upper right panel) and of  $(m_{s},\Delta N_{\rm eff})$ 
for thermally distributed  $\nu_s$ (upper left panel) and for the for 
DW scenario (lower panel) for the  CMB+BAO cosmological data set 
({\it SET 1}).  Black dashed
contours are those of $ -2 \log(\lhood/\lhood^{\rm max})<C$, which
would correspond to nominal 1,2,3 sigma levels.  The red line
denotes the region for which $m_s =10$~eV for thermal $\nu_s$. }\label{fig:set1}
\end{figure}

\begin{figure}
\begin{tabular}{cc}
\multicolumn{2}{c}{\Large CMB+BAO+BICEP2} \\
&  Thermal $\nu_s$ \\
\includegraphics[width=0.5\textwidth,clip=true]{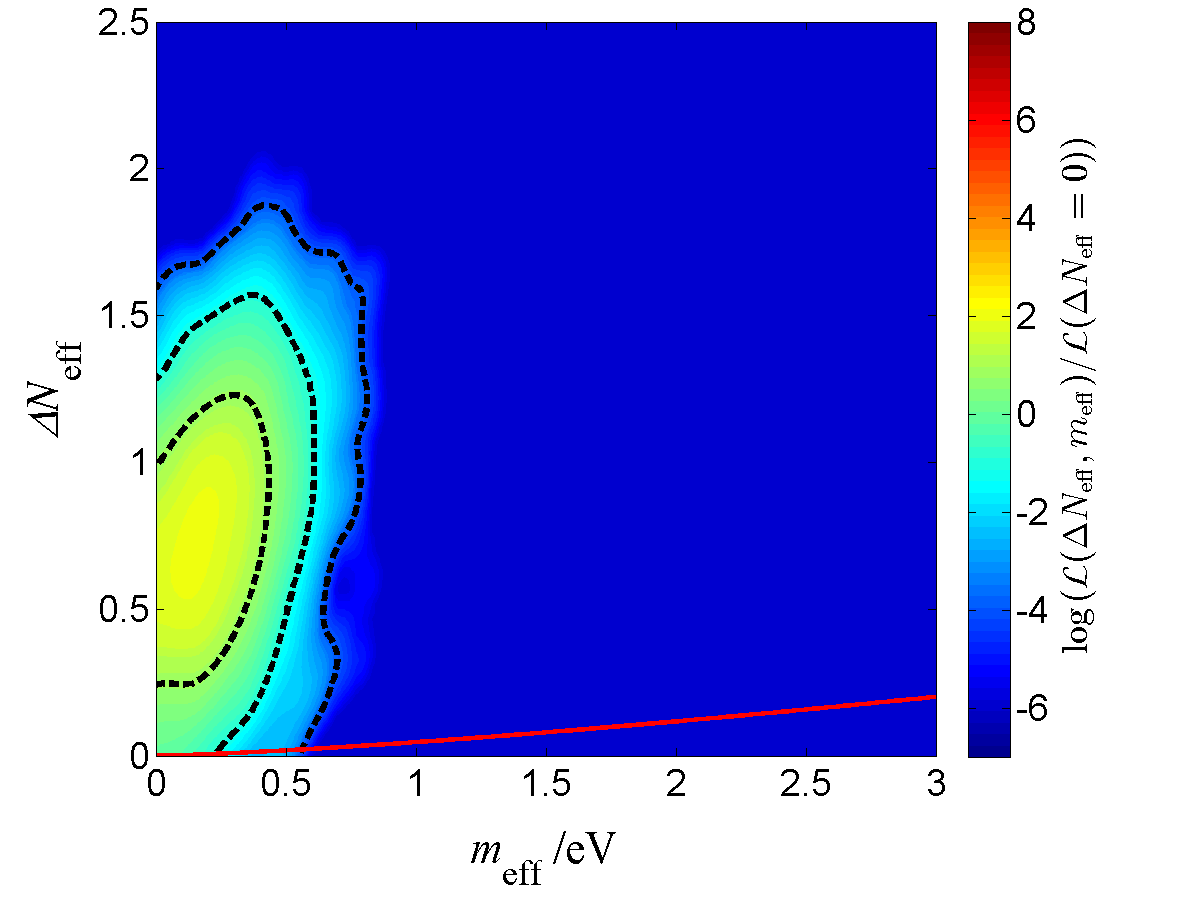} & 
\includegraphics[width=0.5\textwidth,clip=true]{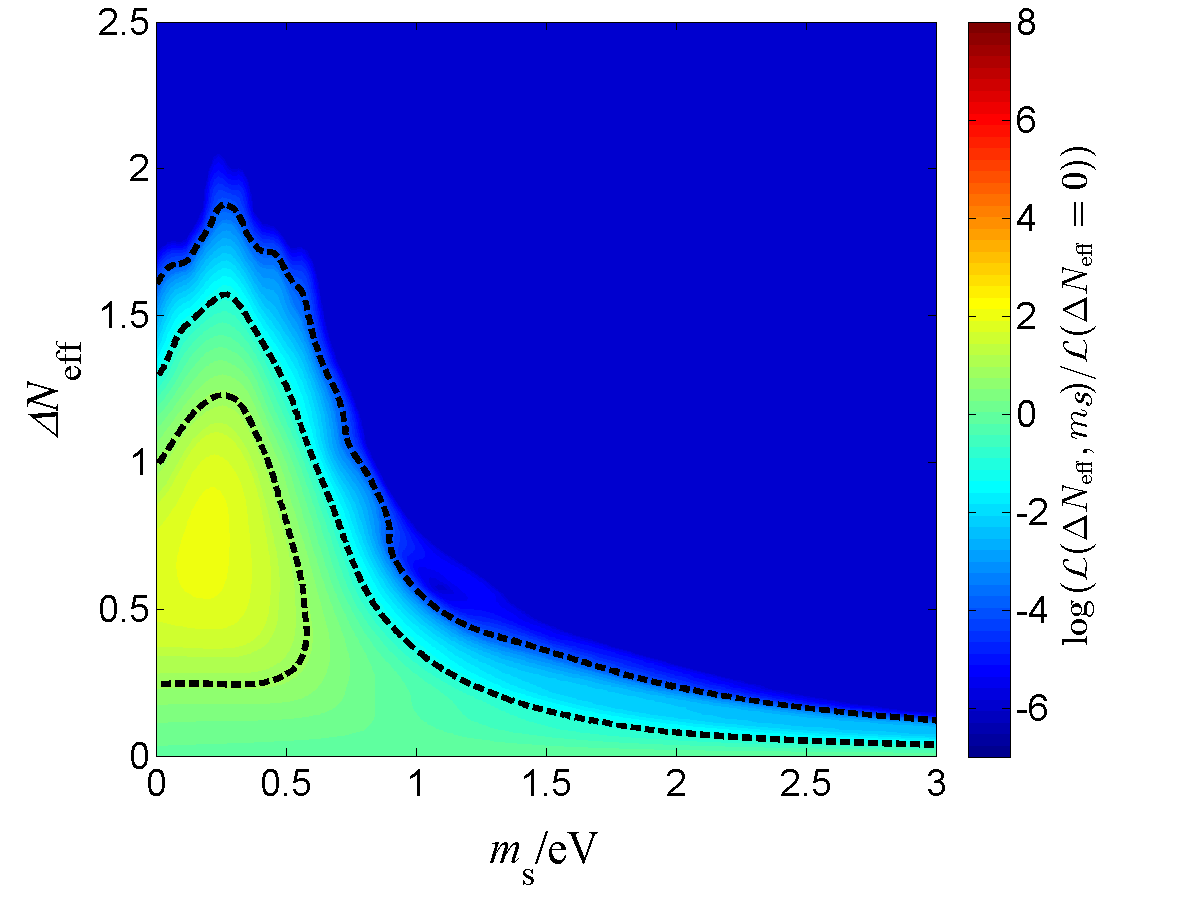} \\
\multicolumn{2}{c}{\Large CMB+BAO+BICEP2+HST+PlaSZ} \\
\includegraphics[width=0.5\textwidth,clip=true]{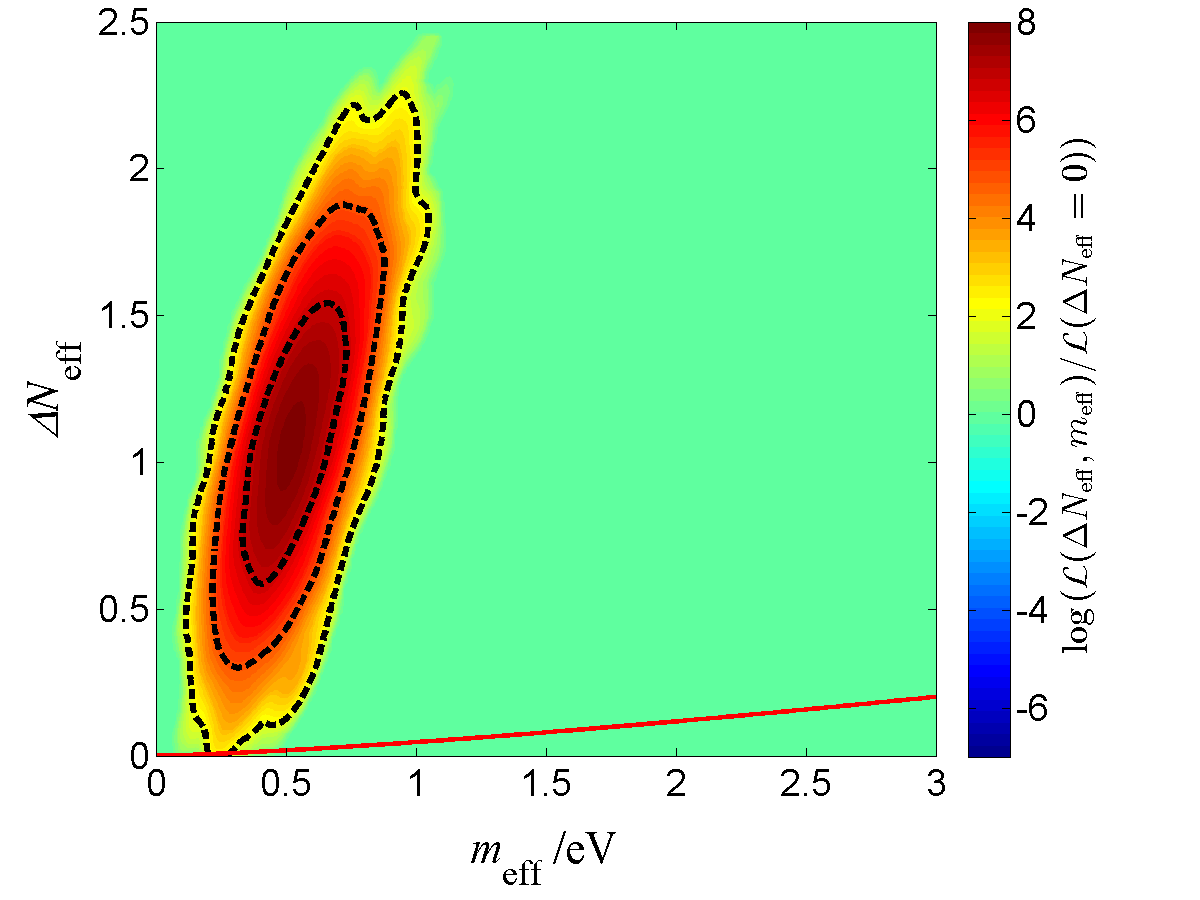} & 
\includegraphics[width=0.5\textwidth,clip=true]{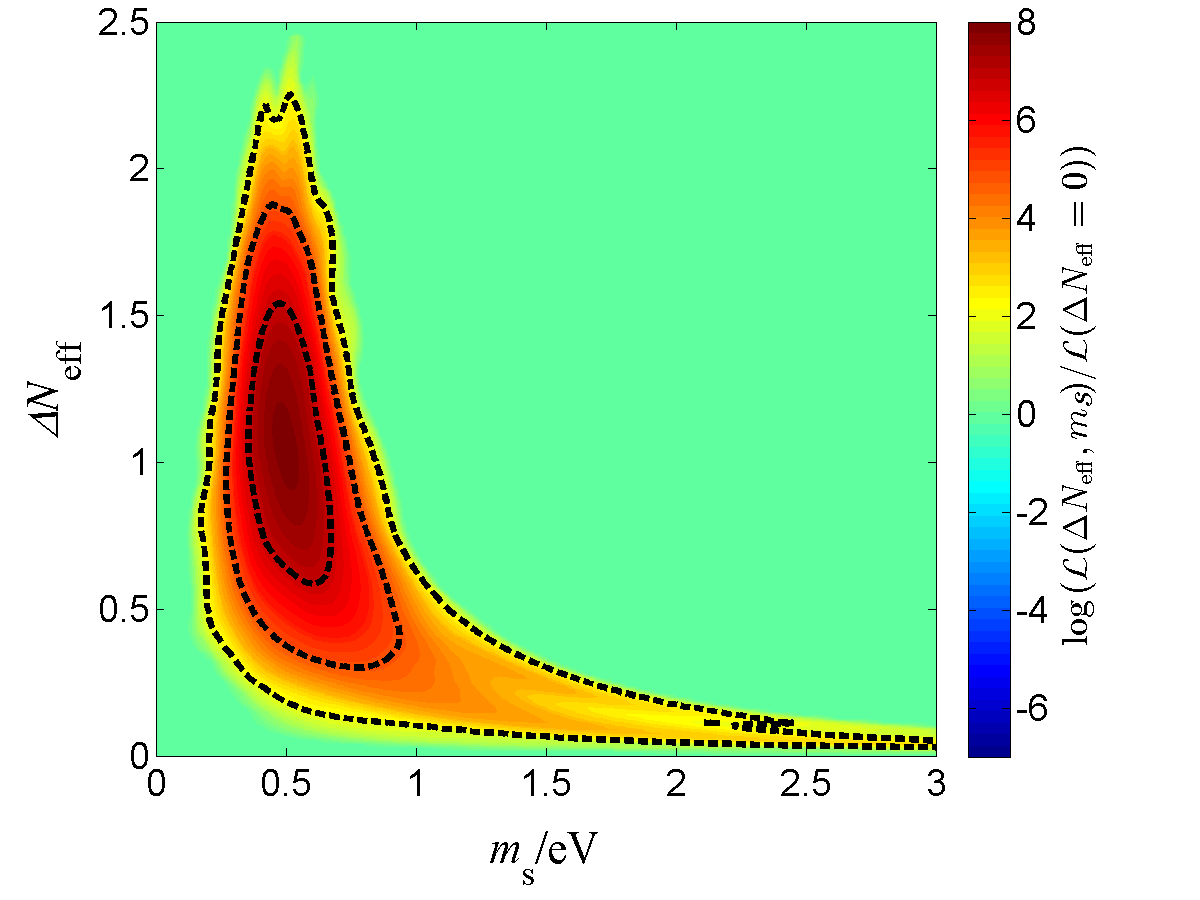} \\
\end{tabular}
\caption{Same two upper panels in  \figref{fig:set1}, 
but for CMB+BAO+BICEP2 cosmological data ({\it SET 2}, upper panels), and
CMB+BAO+BICEP2+HST+PlaSZ cosmological data  ({\it SET 3}, lower panels).}
\label{fig:set2and3}
\end{figure}

The results of our analysis for the three data set combinations
described in \secref{subsec:data} are shown in 
Figs.~\ref{fig:set1} and ~\ref{fig:set2and3}. 

In the the upper left panel of \figref{fig:set1} and in the left
panels of \figref{fig:set2and3} we plot the contours of  
the marginal likelihood $\lhood(N_{\rm eff},m_{\rm eff})$ 
normalized to the value
of $\lhood(\Delta N_{\rm eff}=0)$ (which does not depend on $m_{\rm
eff}$) \footnote{For likelihoods more than eight log-units away from the maximum
value, we extrapolate using a constant value.  
It could be made more accurate by using non-constant
functions such as polynomials but no qualitative change is expected.}.  
The red contour delimits the regions
for which $m_s$ in \eref{eq:msther} exceeds 10 eV 
for which  the sterile states becomes indistinguishable from cold or 
warm dark matter \cite{Ade:2013lta} 
and hence for values of the parameters below the curve, the marginal 
likelihood is not evaluated.  Black
dashed contours are those of $ -2 \log(\lhood/\lhood^{\rm max})<
2.30,6.18,11.83$, nominally corresponding to the $1\sigma$, $2\sigma$,
and $3\sigma$ levels in two dimensions.

In the upper right panel of \figref{fig:set1}  and the right panels
of \figref{fig:set2and3} we show the contours of the marginal
likelihood of $(N_{\rm eff},m_{s})$ for the thermally distributed
sterile neutrinos while in the lower left panel of \figref{fig:set1}
we show the corresponding marginal likelihood in the DW scenario. We
see that for both DW and thermal $\nu_s$ scenarios, $m_s$ becomes increasingly
large for decreasing $\dNeff$, hence the distinctive appearance of
large \qu{flat} regions and weak constraints on $\ms$ for small
$\dNeff$. Also as seen in \figref{fig:set1} the results for 
the DW and the thermal $\nu_s$ scenarios are qualitatively very similar.
Only, since for fixed $m_{\rm eff}$ and 
$\dNeff$, $m_s^{\rm DW}=m_s^{\rm TH} \dNeff^{-1/4}$, for  $\dNeff\leq 1$ 
the likelihood contours in the DW scenario are shifted 
to slightly larger masses in an amount which decreases as $\dNeff$ 
increases.   

The impact of BICEP2 data can be seen by comparing the
upper panels in \figref{fig:set1} and
~\figref{fig:set2and3}.  The addition of BICEP2 data gives a
preference for large values of the tensor-to-scalar ratio $r$ and, due
to its correlation with $\dNeff$, and therefore leads to the the shift 
to slightly larger values of
$\dNeff$ observed in upper panels in ~\figref{fig:set2and3}.
Since $r$ is now better constrained, also the preferred
region becomes slightly smaller.  

The effects of adding HST+PlaSZ in the analysis are displayed in the lower
panels in \figref{fig:set2and3} where we see a shift to larger 
values of both $\dNeff$ and $\meff$. This is so because the constraints on
$\sigma_8$ of the PlaSZ measurement, which are in tension with the
other experiments within this model, can be somewhat alleviated by an
increase in  $\meff$, while the inclusion of HST yields an
increase $\dNeff$.  

All these results are in qualitative agreement with those in the analyses
in Refs.~\cite{Archidiacono:2014apa,
Dvorkin:2014lea,Zhang:2014dxk,Zhang:2014nta,Wu:2014qxa,
Leistedt:2014sia,Giusarma:2014zza}. However we notice that by showing
the marginal likelihood we are explicitly not assigning any priors to
the sterile parameters.  This is an advantage in the absence of a
physical motivation for them,  
especially $\ms$, since the data at hand
is expected to leave a large prior dependence.  For example, if one
used a prior on $\ms$ which is uniform in $\log \ms$ instead of in
$\ms$, as in general the likelihood is non-negligible for a vanishing
sterile mass, the derived Bayesian constraints on $\meff$ and $\dNeff$
would be very different.  In principle one could embark on an
extensive prior sensitivity analysis, but in this work we will instead
focus on analyses for which the results have little or no prior
dependence.

\begin{figure}
\begin{tabular}{cc}
CMB+BAO & CMB+BAO+BICEP2\\
\includegraphics[width=0.5\textwidth,clip=true]{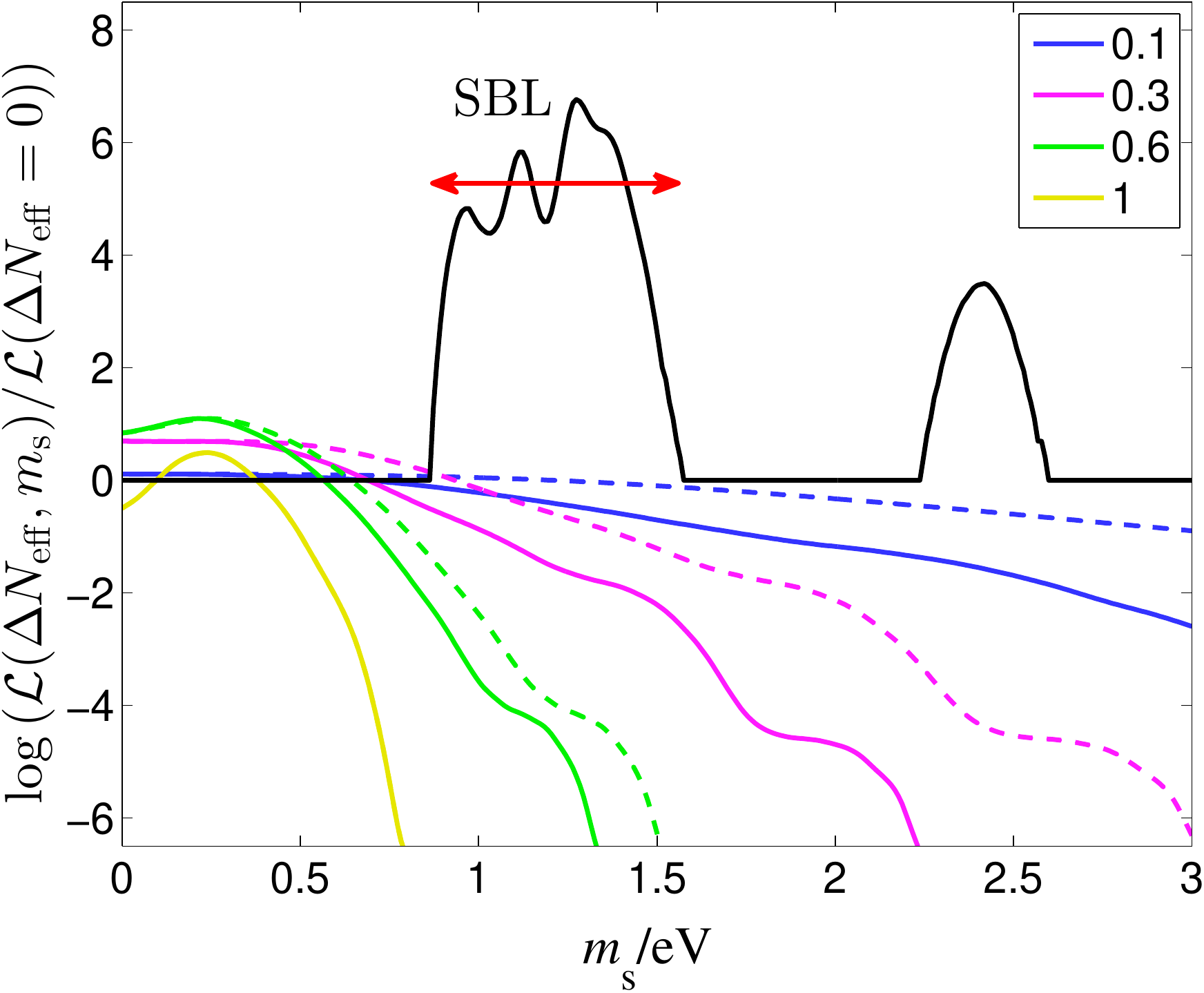} &
\includegraphics[width=0.5\textwidth,clip=true]{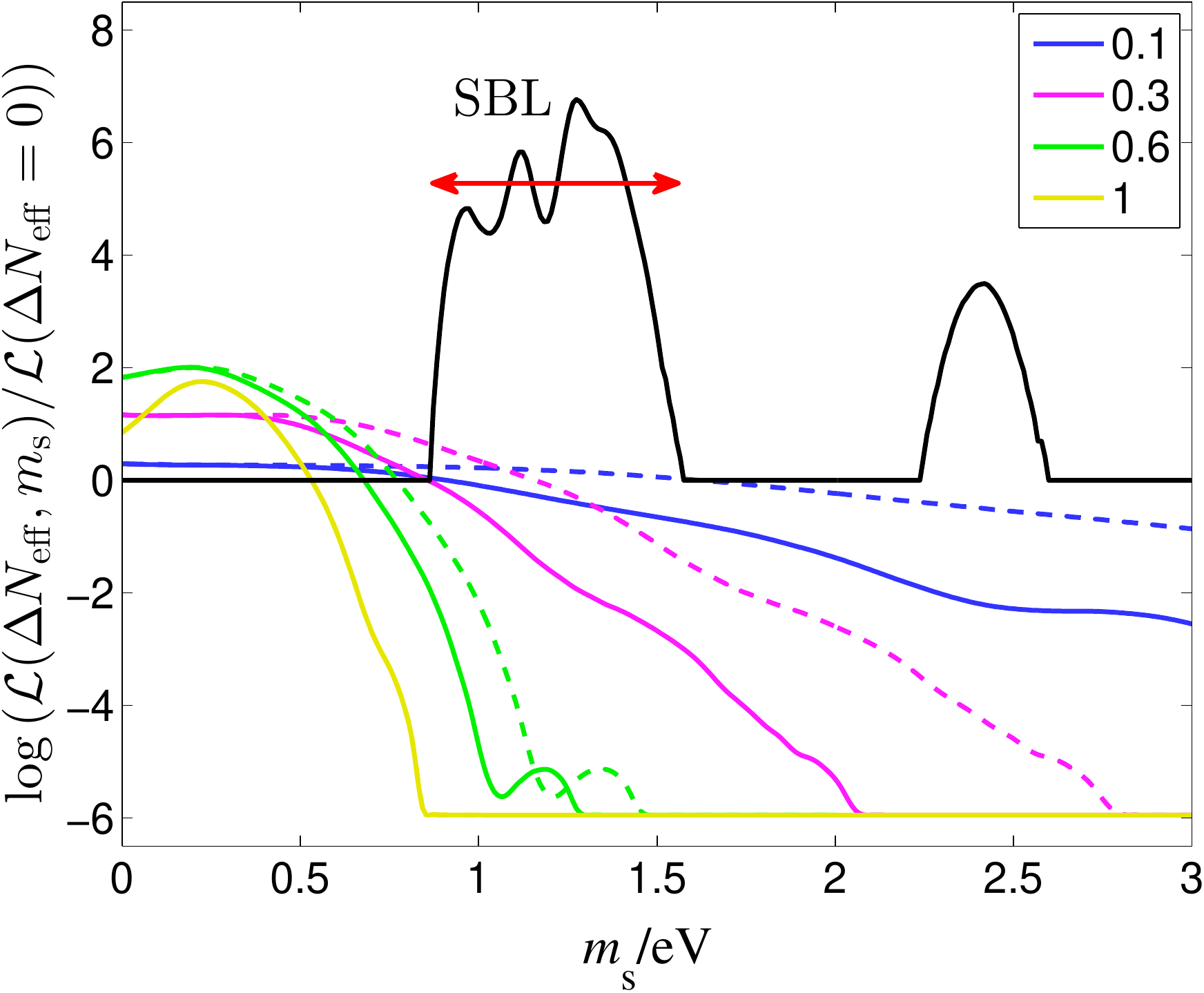} \\ 
 \multicolumn{2}{c}{CMB+BAO+BICEP2+HST+PlaSZ}\\
\multicolumn{2}{c}{\includegraphics[width=0.5\textwidth,clip=true]{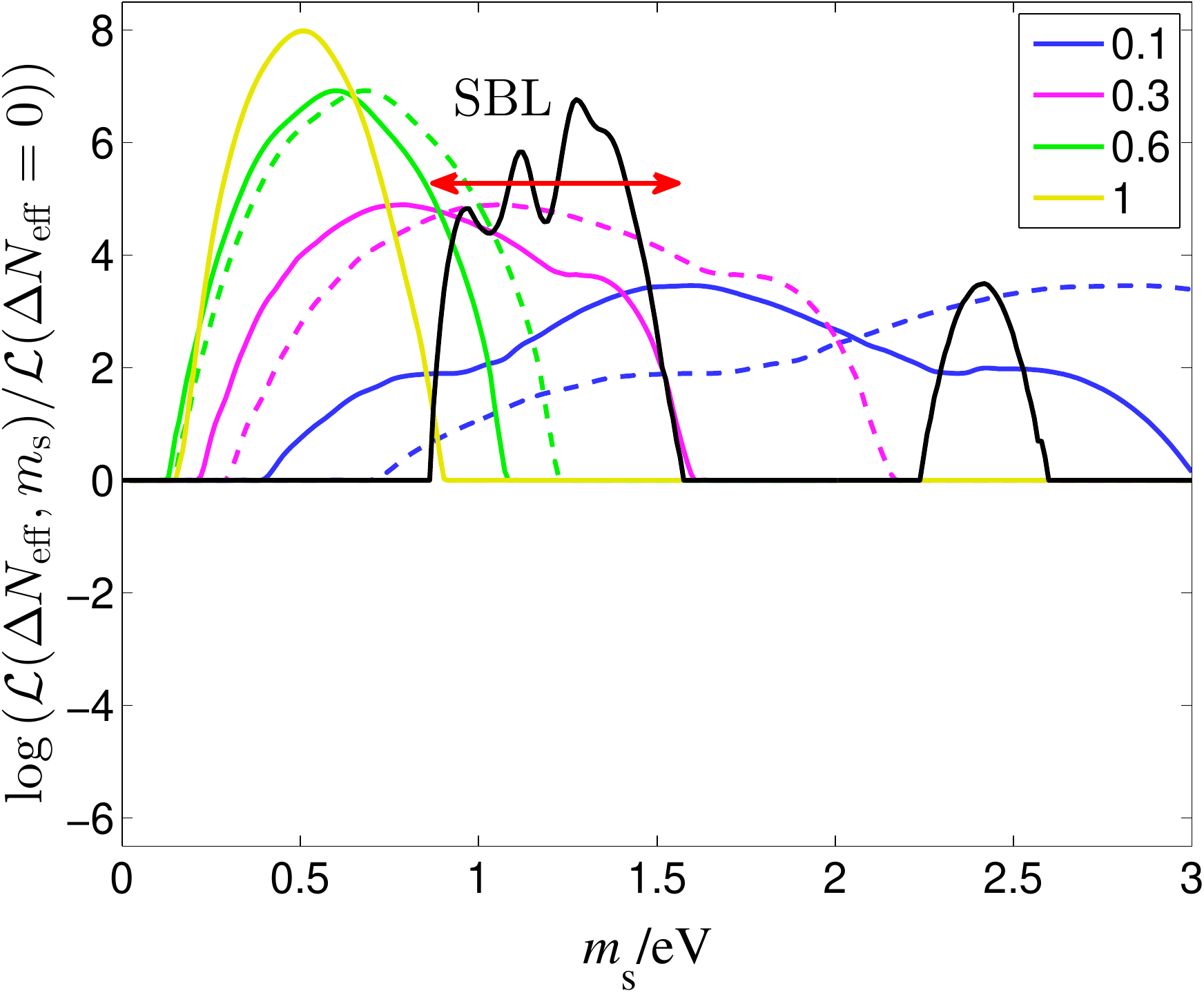}  }
\end{tabular}
\caption{Marginal likelihoods as function of $m_{s}$ for fixed $\Delta
N_{\rm eff}$ ($\Delta N_{\rm eff}= 0.1, 0.3, 0.6, 1$) and 
for thermal $\nu_s$ (solid lines) and for the DW scenario (dashed lines). We
show the results for the three cosmological data sets used as labeled
in the figure. In all panels we also include the marginal likelihood  
for the SBL analysis in the 3+1 scenario (marginalized with respect
to the lighter neutrino masses and all mixings) as 
given in Fig.1 in \refcite{Kristiansen:2013mza}. We denote by the
red arrow the width and height of the box used to define 
``box SBL likelihood'' (see text for details).}
\label{fig:sterilenu_1D_TH_log}
\end{figure}

Finally, in order to better illustrate how the constraints depend on the
sterile mass we plot in \figref{fig:sterilenu_1D_TH_log}, the slices of
the marginal likelihood as a function of $m_{s}$ for fixed $\Delta
N_{\rm eff}$.  Also shown in the figure is the marginal likelihood  
for the SBL analysis in the 3+1 scenario (marginalized with respect
to the lighter neutrino masses and all mixings) as 
given in Fig.1 in \refcite{Kristiansen:2013mza}.

%---------------------------------------------------------------------
\section{Test on sterile neutrinos models}
\label{sec:sterile}
%---------------------------------------------------------------------
In this section we perform the statistical tests on the 
3+1 scenarios invoked to explain the SBL anomalies using the results of the
cosmological analysis presented in the previous section. 
In doing so we are going to assume that the contribution of the 
sterile neutrino to the cosmological observables is independent of the 
active-sterile neutrino mixing (see discussion around \eref{eq:FNT}).
Under this assumption our tests will require  the marginalized likelihood 
of the sterile neutrino mass $m_s$ obtained from the analysis of the 
SBL anomalies in the 3+1 scenario (marginalized
with respect to all other oscillation parameters in the scenario),  
and the marginalized likelihoods of $(m_{s},\Delta N_{\rm eff})$ 
from the cosmological analysis previously presented.  

Concerning the SBL likelihood, we consider two different functions that
can be interpreted as the two limiting cases. In the first, we
consider the precise SBL likelihood as given in Fig. 1 in
\refcite{Kristiansen:2013mza} which we reproduce in
\figref{fig:sterilenu_1D_TH_log} (in this case below roughly seven
log-units from the maximum value, we set the SBL likelihood to a
constant value).  We label this case in the following as ``full SBL
likelihood''.  In the second case, we approximate the SBL likelihood
as a top-hat shaped likelihood which is constant and non-zero between
0.86 eV and 1.57 eV and zero otherwise (which we illustrate by the
arrow in \figref{fig:sterilenu_1D_TH_log}).  This is what we label in
the following as ``box SBL likelihood''.

\subsection{Sterile neutrinos vs none}

The first question we want to address is whether the current  
data shows evidence of the existence of sterile neutrinos, and how strong this
evidence is. Generically in Bayesian analyses this takes the form 
of model comparison between a model without sterile neutrinos and a
model with sterile neutrinos using some posterior odds, 
\eref{eq:post_ratio}.  
There are several ways to go about answering this question. 
In here we are interested in testing what cosmology has to say on this
comparison for the sterile models invoked to explain the SBL anomalies. 
This is, in this case the first model $H_0$ is defined as a model with no
sterile neutrino,  which implies a cosmological model with
$\dNeff = 0$.  And the other model, $H_1$ is taken to include one sterile
neutrino of mass $m_{\rm s}$ {\sl as required to accommodate the SBL
anomalies}  and which contributes to relativistic
energy density in the early Universe as some fix $\dNeff$. 

In this case we can define a posterior odds as:
\bea 
\mathcal{O}\notag & =&
\frac{\Pr(\Dc|\Dsbl,H_1)}{\Pr(\Dc|\Dsbl,H_0)}
\frac{\Pr(H_1|\Dsbl)}{\Pr(H_0|\Dsbl)}
=\\
&=&\frac{\Pr(\Dc|\Dsbl,H_1)}{\Pr(\Dc|\Dsbl,H_0)}
\frac{\Pr(\Dsbl|H_1)}{\Pr(\Dsbl|H_0)}
\frac{\Pr(H_1)}{\Pr(H_0)}\label{eq:O_upd} \\ 
&\equiv &\mathcal{B}^\upd_{10} 
\mathcal{B}^{\rm SBL}_{10} 
\frac{\Pr(H_1)}{\Pr(H_0)}
\,, \notag\eea 
with
\bea \mathcal{B}^{\rm SBL}_{10}& \equiv & \frac{\Pr(\Dsbl|H_1)}{\Pr(\Dsbl|H_0)}\eea 
and 
\bea \mathcal{B}^\upd_{10} & \equiv & 
\frac{\Pr(\Dc|\Dsbl,H_1)}{\Pr(\Dc|\Dsbl,H_0)} = 
\frac{\Pr(\Dc|\Dsbl,H_1)}{\Pr(\Dc|H_0)} = 
\frac{\Pr(\Dc,\Dsbl|H_1)}{\Pr(\Dc|H_0) \Pr(\Dsbl|H_1)} \,,
\label{eq:Bupd}
\eea 
where in the last line we have used that SBL is not sensitive to any 
parameters effecting cosmology once we assume that there are no sterile 
neutrinos and hence  $\Dsbl$ and $\Dc$ are  independent under $H_0$.
The quantity $\mathcal{B}^\upd_{10}$ quantifies how much better the
prediction of cosmological data assuming sterile neutrino and SBL data
is than the prediction assuming no sterile neutrinos. 

Now, the model $H_1$ is inherently more complex than $H_0$ since it
contains additional parameters (sterile mass and mixings), and
this is as usual compensated for in a Bayesian analysis. In the first
row of \eref{eq:O_upd} this is contained in the last factor of
$\Pr(H_1|\Dsbl)/\Pr(H_0|\Dsbl)=\mathcal{B}^{\rm SBL}_{10} 
\frac{\Pr(H_1)}{\Pr(H_0)}$. Now, the best-fit of the SBL data
is significantly better if you have a sterile neutrino,   
but because of the added complexity it might not be totally 
unreasonable to have  $\Pr(H_1|\Dsbl)/\Pr(H_0|\Dsbl) = \mathcal{O}(1)$.

In any case, it is the first factor in \eref{eq:O_upd} 
the factor by which
the cosmological data have \emph{updated} the SBL-only posterior odds
to the final SBL+cosmology odds and which we will be using in our
quantification. 
Furthermore, under
the additional assumption that $\dNeff$ is unconstrained by SBL data,
$\Pr(\Dsbl|H_1)$ does not depend on $\dNeff$, and hence $\Bupd$ is in
fact simply proportional to the combined marginal likelihood of
$\dNeff$, normalized such that $\Bupd = 1$ for $\dNeff = 0$. 

Before discussing the results, let us mention that here, as in any Bayesian 
analysis, the results are in principle always prior
dependent, and we should consider how large this dependence is in
practice. First, as discussed in \secref{sec:cosmo}, the marginal
likelihood depends on the priors on the cosmological nuisance
parameters, but this dependence is expected to be small (except
possibly for $r$). More significantly, there is the dependence of the
total Bayes factor and odds on the prior on $m_s$, even when it is
well constrained by the combined data set.  
However the value of $\Bupd$ does not strongly depend on the the shape 
of the prior nor its upper limit. This is so because the well-constraining 
SBL data is used to update the prior to a posterior 
(which is rather insensitive to the prior) which
is then used to analyze the cosmological data.  We use a uniform prior
between $0$ and $ 10 \eV$ as the nominal upper limit, since, as
described in \refcite{Ade:2013lta}, this roughly defines the region
where (for the CMB) the particles are distinct from cold or warm dark
matter. 

\begin{figure}
\begin{tabular}{cc}
 \multicolumn{2}{c}{CMB+BAO} \\
\includegraphics[width=0.5\textwidth,clip=true]{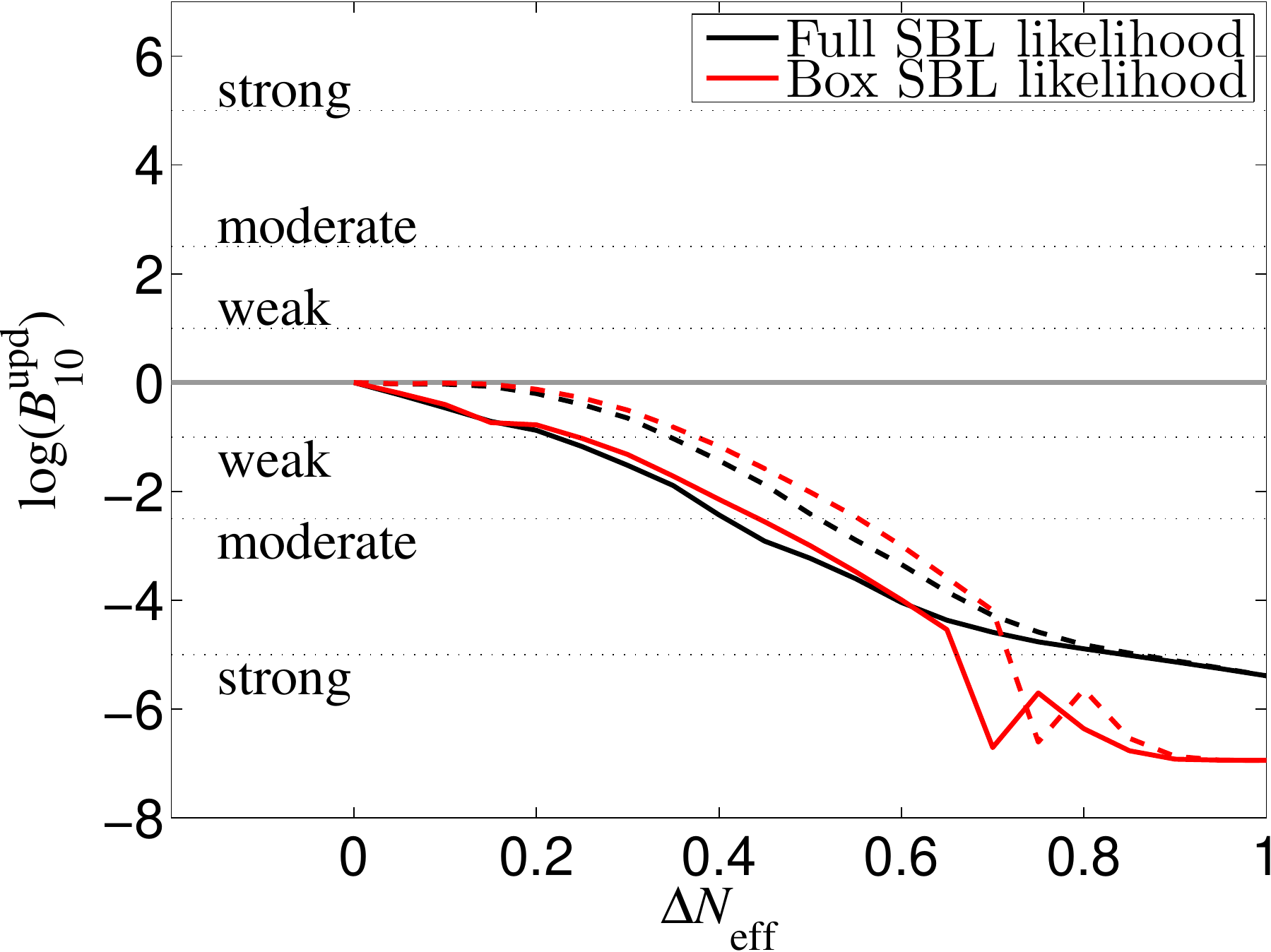}&
\includegraphics[width=0.45\textwidth,clip=true, height=0.38\textwidth]{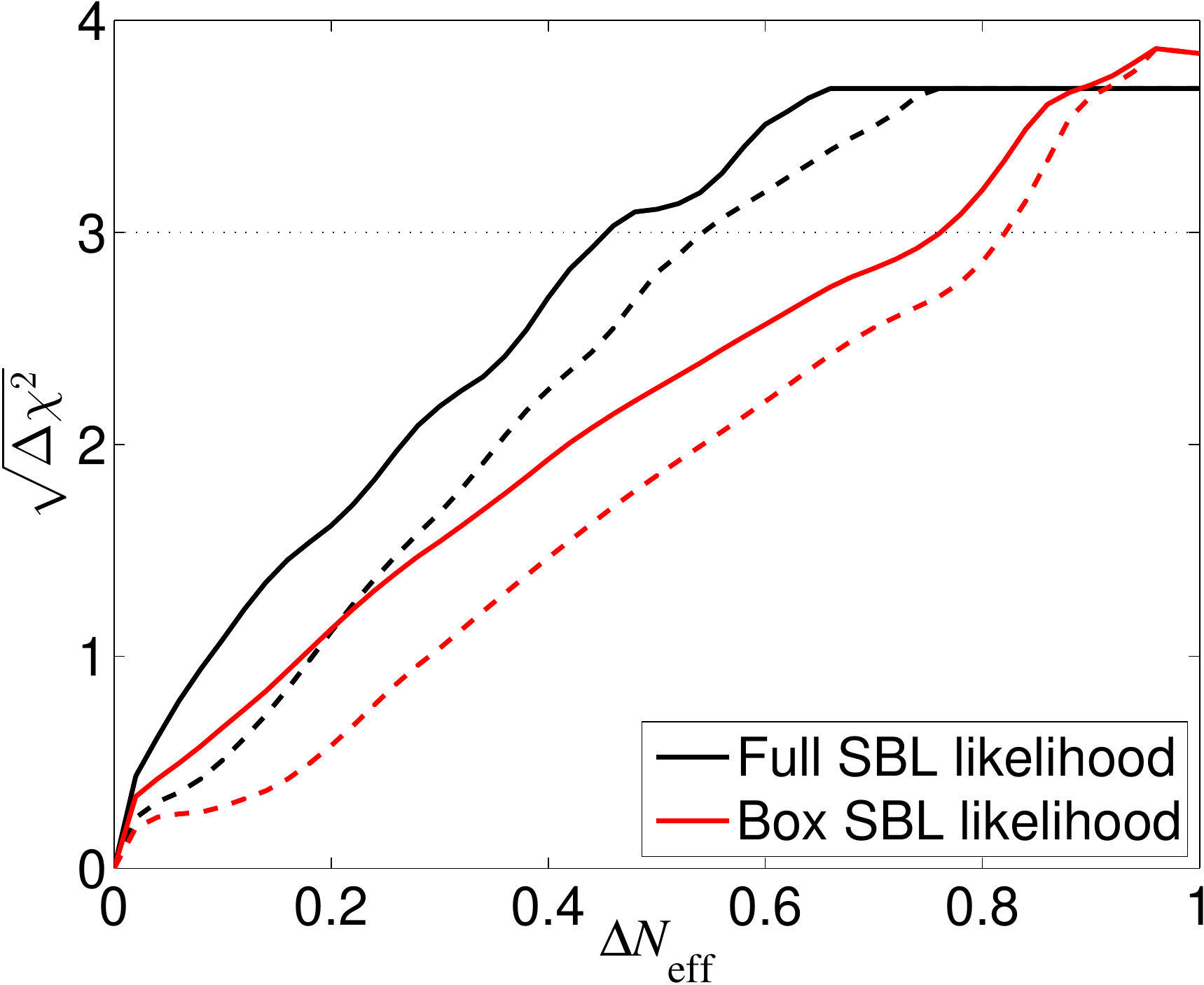} \\
 \multicolumn{2}{c}{CMB+BAO+BICEP2} \\
\includegraphics[width=0.5\textwidth,clip=true]{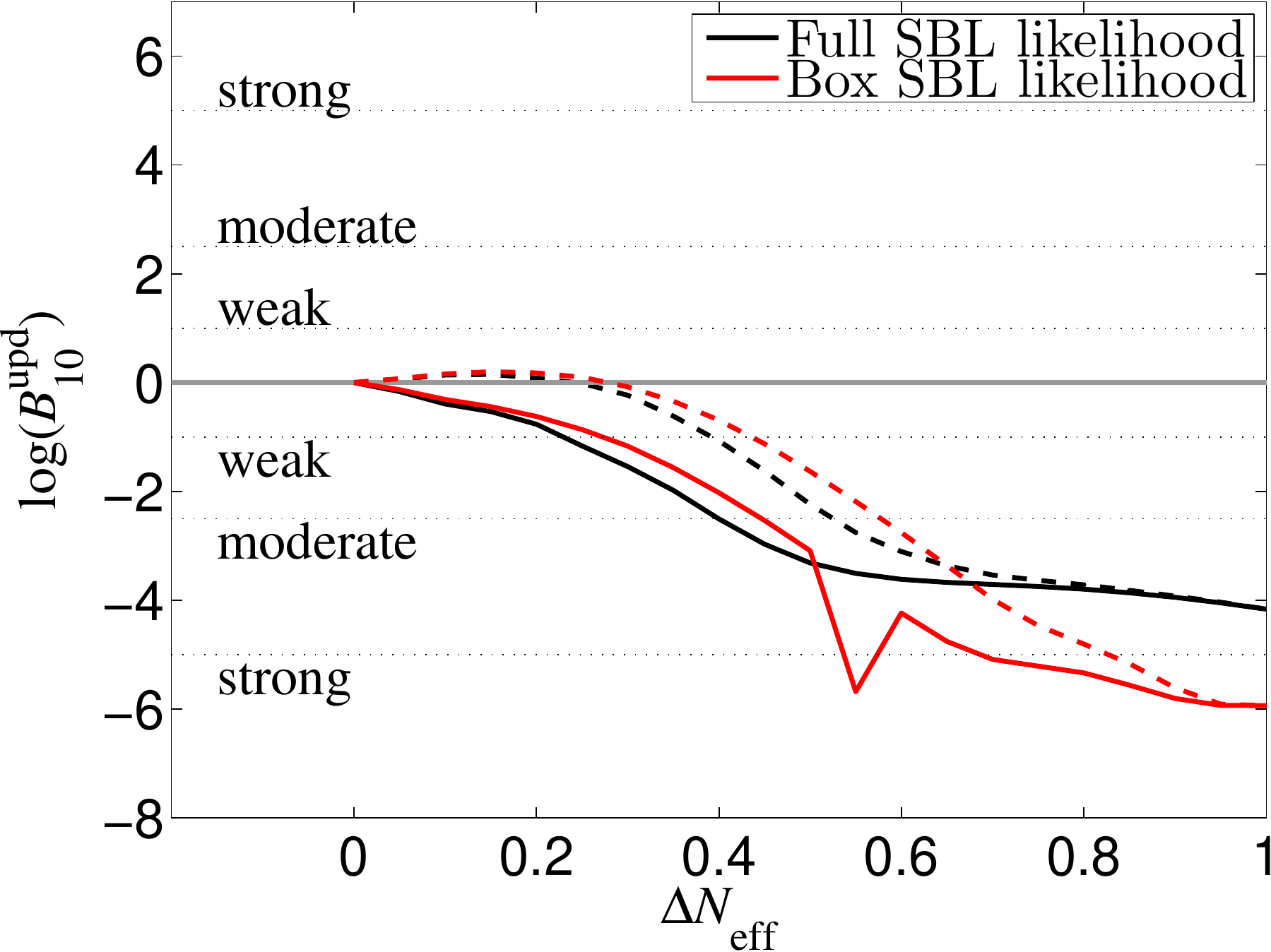}&
\includegraphics[width=0.45\textwidth,clip=true, height=0.38\textwidth]{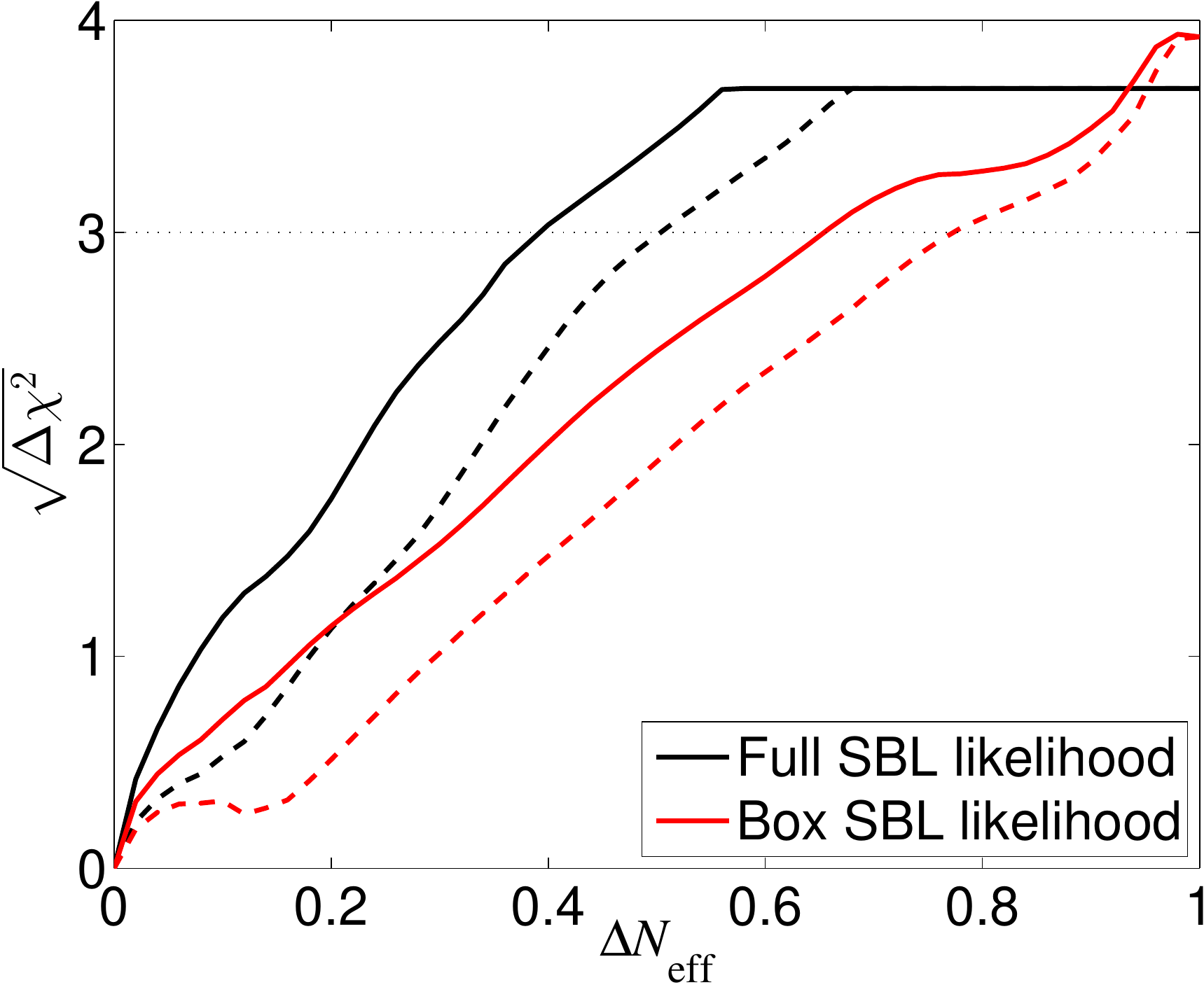} \\
 \multicolumn{2}{c}{CMB+BAO+BICEP2+HST+PlaSZ} \\
\includegraphics[width=0.5\textwidth,clip]{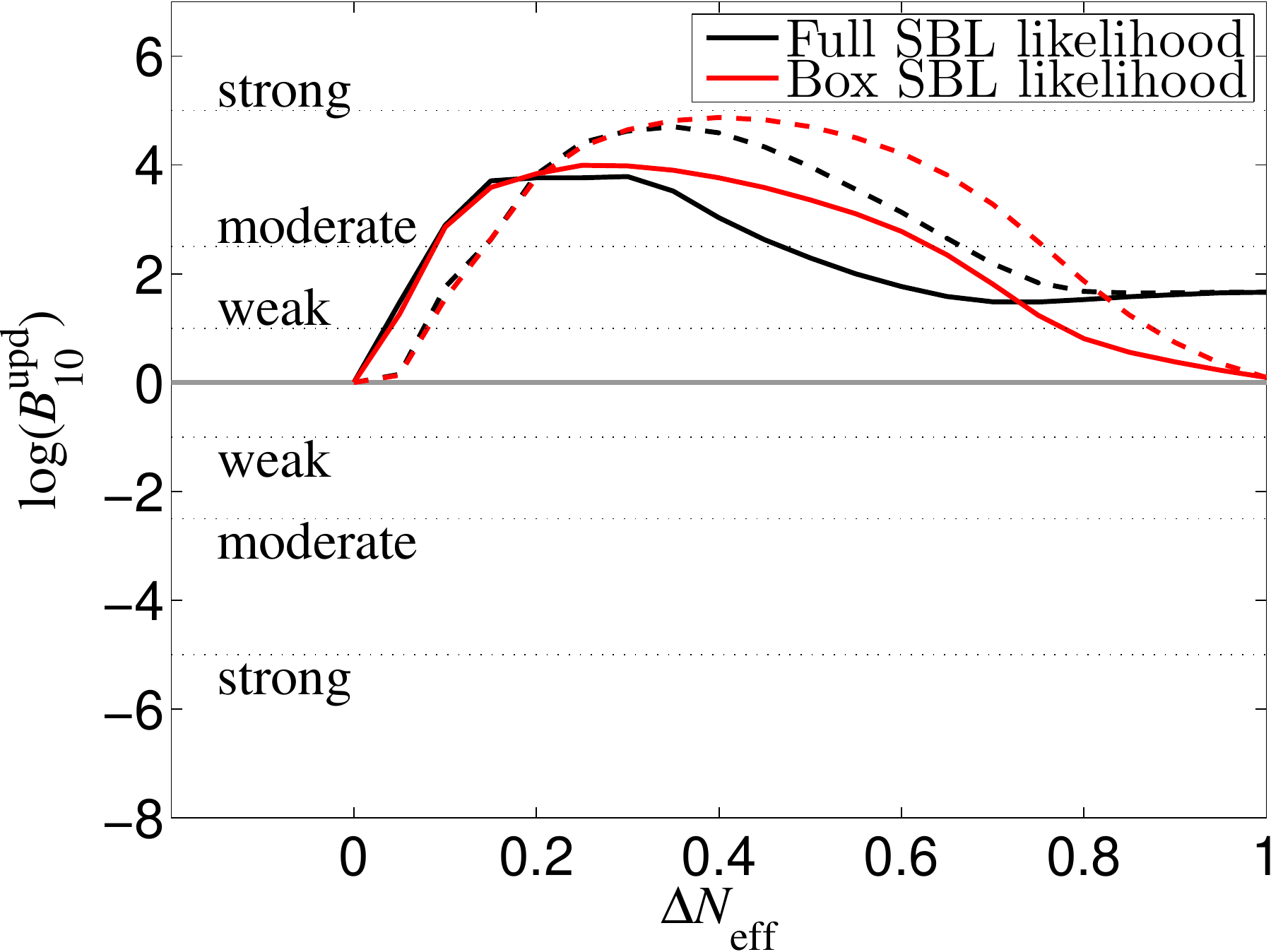} &
\includegraphics[width=0.45\textwidth,clip=true, height=0.38\textwidth]{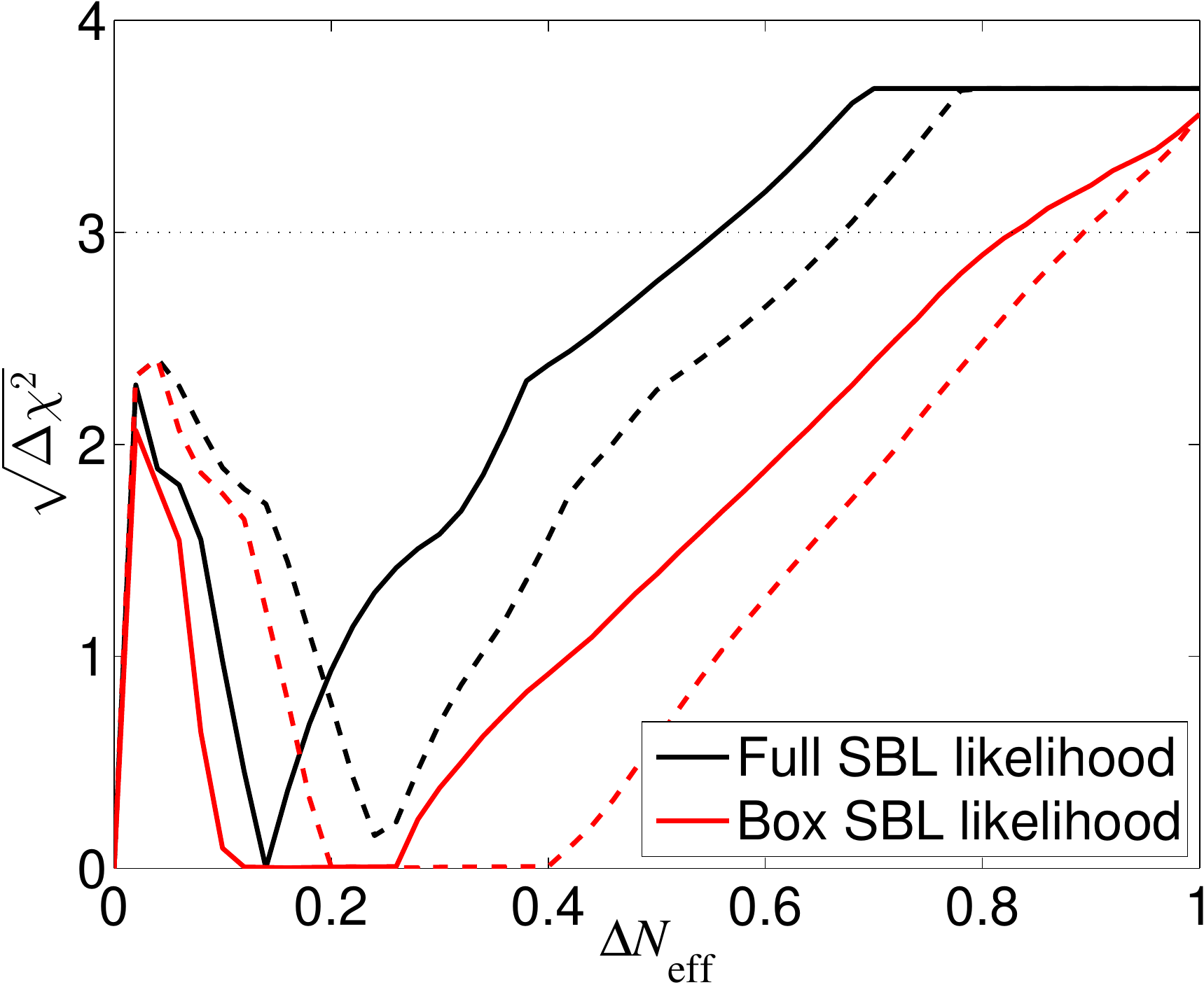} 
\end{tabular}
\caption{Left panels:Logarithm of the Bayes factor $\Bupd$ 
as a function of $\Delta N_{\rm
eff}$. Right panels: Consistency of mass constraints. 
In all panels the results are shown for  thermal $\nu_s$  (solid lines) 
and for the DW scenario
(dashed lines) and for the SBL full (black) and Box likelihood (red).
We show the results for the three cosmological data
sets considered:  CMB+BAO (upper row), CMB+BAO+BICEP2 (middle row),
CMB+BAO+BICEP2+HST+PlaSZ (lower row).}
\label{fig:Bupf-chi2}
\end{figure}

The results are shown in the left panel of \figref{fig:Bupf-chi2}. 
We see that for the CMB+BAO and CMB+BAO+BICEP2 data sets, $\Bupd$ 
decreases quite steadily with $\dNeff$. These cosmological data hence 
disfavour models  with sterile neutrinos required to explain the SBL anomalies 
over the model without sterile neutrinos independently of how much 
the contribution of the sterile neutrino to the energy density is suppressed
with respect to the fully thermalized expectation. 
We also see that the addition of the BICEP2 data has a small impact on 
these conclusions. 

For the CMB+BAO+BICEP2+HST+PlaSZ data set, there is, on the contrary, 
a significant peak for intermediate values of $\dNeff$.   
The $\Lambda$CDM$+r$ model is
significantly disfavoured by this combination of cosmological data, 
while increasing $\dNeff$ increases the cosmological likelihood 
for SBL-compatible masses. 
However, further increasing $\dNeff$, cosmology requires a too small mass, so
the $\Bupd$ decrease again. Notice also that in fact the $\Lambda$CDM$+r$ is so
disfavored that it is in the region where the (too large) constant
extrapolation is used. Hence, the exact $\Bupd$ is expected to be
even larger.

\subsection{Consistency of parameter constraints}
In addition to comparing the models with and without sterile neutrinos 
we now address the question of the consistency  of the parameter constraints 
from the different data sets within the 3+1 model.
A Bayesian test was formulated in \cite{Marshall:2004zd}, in which a
model where both data sets are fitted by the same physical parameters
is compared with a model in which each data set uses their own
parameters. However, as is often the case in model comparison, the
result can depend crucially on the prior on the parameters, in our
case $m_s$ in particular. Since we cannot motivate using a specific
shape or limits on the prior, we instead use the corresponding
$\chi^2$-test (these were also compared in
\cite{Marshall:2004zd}). Although not rigorous as the Bayesian test,
it has the clear advantage that it is prior-independent.

In particular, if we want to test how inconsistent the constraints on the 
sterile mass from the different data sets are once a certain 
$\dNeff$ is assumed,  {\it i.e.}, without considering how
favoured or disfavoured that $\dNeff$ is by the cosmological data, we 
should evaluate
\begin{equation}
\Delta \chi^2 (\dNeff) = \hat{\chi}^2_{\rm comb}(\dNeff) - \hat{\chi}^2_{\rm SBL} -  \hat{\chi}^2_{\rm cosmo}(\dNeff) \; , , 
\label{eq:chi2comp}
\end{equation}
where the hat denotes the value at the best fit {\it i.e.}, optimized
over $m_s$ ($ \chi^2_{\rm SBL}$ does not depend on $\dNeff$). 

The results of this test are shown on the right-hand side of 
\figref{fig:Bupf-chi2}.
As expected, the results for CMB+BAO and CMB+BAO+BICEP2 are quite
similar and  show an steady increase of the inconsistency with 
$\dNeff$. Also comparing the DW and thermal scenarios, we see that in general
to obtain the same $\Delta \chi^2$ larger values of $\dNeff$  are
required in the DW scenario. This is so because, as explained before,  
in the DW scenario the preferred masses are shifted to larger values 
by an amount which increases  as $\dNeff<1$ decreases. 
As for the results for the box and full  
SBL likelihoods, we notice that, typically, using the full SBL 
likelihood gives a larger  inconsistency. This is because the box likelihood 
is constant over a wide 
range of $m_s$, and over this wide range the cosmological likelihood 
typically varies significantly. The combined $\chi^2$ can then easily 
be reduced by finding the best fit within this range.

For the CMB+BAO+BICEP2+HST+PlaSZ combination, one observes in each curve 
a peak for small value of $\dNeff$, although it does not reach the level of
$2.5 \sigma$. These peaks are due to the fact that
(cf. \figref{fig:set2and3}) cosmology prefers large physical
masses, too large to fit the SBL data. As $\dNeff$ increases, the
mass constraints become compatible, but as $\dNeff$ continues to
decrease, cosmology requires the masses to be smaller than those which
can fit the SBL data and the inconsistency increases.

Again, we stress that this consistency test is in principle not
affected by how favoured or disfavoured the considered value of
$\dNeff$ is, but instead consider that value to be \qu{true}, and then
test the compatibility of the constraints on the mass.  For example,
comparing the left and right panels in the last raw of
\figref{fig:Bupf-chi2} we see that for the CMB+BAO+BICEP2+HST+PlaSZ
data set, even though from the right panel we read that the mass
ranges required for cosmology and SBL become highly incompatible for
$\dNeff$ close to one, from the left hand side we see that these large
values of $\dNeff$ are not particularly disfavoured compared to small
$\dNeff$, for which the mass constraints are compatible. So what we
see is that large $\dNeff$ is disfavoured because the sterile mass
required by SBL and cosmology are incompatible and small $\dNeff$ is
disfavoured because it is so in the cosmological (and consequently in
the combined) analysis.  So small and large $\dNeff$ have comparable
Bayes factors, but this is only because they are both disfavoured.

%---------------------------------------------------------------------
\section{Summary}\label{sec:conclusions} 
%---------------------------------------------------------------------

In this paper we have revisited the question of the information which
cosmology provides on the scenarios with {\cal O}(eV) mass 
sterile neutrinos invoked to explain the SBL anomalies 
(\eref{eq:parsbl})  using Bayesian statistical tests and study how
the results depend on the inclusion of the recently CMB polarization
results of BICEP2 and on the inclusion of local measurements which 
show some tension with the Planck and LSS-BAO results when analyzed
in the framework of the $\Lambda$CDM scenario. 

In order to do so we have first performed an analysis of three characteristic 
sets of  cosmological data in $\Lambda$CDM$+r+\nu_s$ 
cosmologies as described in \secref{sec:cosmo}. The result of 
our analysis is presented in Figs.~\ref{fig:set1} and ~\ref{fig:set2and3}
in the form of  marginalized cosmological likelihoods in terms of the two 
relevant parameters, the sterile neutrino mass $m_s$ and its contribution
to the energy density of the early Universe $\dNeff$. The results clearly
indicate that as long as the HST and SZ cluster data from Planck are not
included, cosmological data favours the sterile neutrino mass 
$m_s$  clearly well below $eV$  unless its contribution to the 
energy density is suppressed with respect to the expected from 
a fully thermalized sterile neutrino. The inclusion of the BICEP2 data
does not substantially affect this conclusion. 
Conversely including these 
HST and SZ cluster data higher sterile masses 
become favoured. 

With these results, we have performed in \secref{sec:sterile} two 
statistical test on their (in)compatibility with the corresponding
likelihood derived from the analysis of the SBL results as given in 
Ref.~\cite{Kristiansen:2013mza}.  
In the first test we have asked ourselves whether cosmology favours
or disfavours the 3+1 sterile models which explain the SBL results
over a model without sterile neutrinos. In order to do so 
we have constructed the Bayes factor defined in 
\eref{eq:Bupd} which gives the factor by which
the cosmological data \emph{updates} the SBL-only posterior odds
of the 3+1 vs 3+0 model to the final SBL+cosmology odds and we have studied
its behaviour as a function of $\dNeff$. The results of this test, 
shown in \figref{fig:Bupf-chi2} implies that as long as the 
HST and SZ cluster data from Plank is not included, the cosmological analysis
disfavour the 3+1 model with respect to 3+0. The inclusion of these
cosmological data however favours the 3+1 model for an intermediate range
of $\dNeff$. We summarize in \tabref{tab:favranges} the 
ranges of
$\dNeff$ for which we find the evidence against or in favour of the 3+1 model 
compared
to the model with only the 3 active neutrinos 
to  be weak, moderate or  strong from these analyses. 
\begin{table}
{\small
\begin{tabular}{|c|c||c|c|l@{\hskip 0.08cm}l|}
\hline
Evidence & Scenario
& SET 1 & SET 2& \multicolumn{2}{c|}{SET 3} \\ 
&/SBL Likel& 3+1 Disfavoured & 3+1 Disfavoured & 
\multicolumn{2}{c|}{3+1 Favoured} \\ \hline
    & TH/FULL & $[0.22,  0.40]$ & 
$[0.23,  0.40]$ & $[0.03,  0.08]$\;$\oplus$
&  
$\geq 0.47$ \\
Weak & TH/BOX  & $[0.25,  0.44]$ & $[0.27,  0.45]$ 
& $[0.04,  0.09]$\;$\oplus$&  $[0.63,  0.78]$ \\
    & DW/FULL & $[0.35,  0.51]$ & $[0.39,  0.52]$ 
& $[0.08,  0.14]$\;$\oplus$ & $\geq 0.67$ \\
    & DW/BOX & $[0.38,  0.55]$ & $[0.44,  0.58]$ 
& $[0.08,  0.14]$\;$\oplus$& $[0.76,  0.87]$ \\\hline
    & TH/FULL & $[0.40,  0.86]$ & 
$\geq 0.40$ & 
\multicolumn{2}{c|}{$[0.08 ,  0.47]$}  \\
Moderate & TH/BOX  & $[0.44,  0.66]$ 
& $[0.45 ,  0.53]$ \;$\oplus$ \;$[0.57,  0.68]$ & 
\multicolumn{2}{c|}
{$[0.09 ,  0.63]$} \\
    & DW/FULL & $[0.51,  0.86]$ &  $\geq 0.52$  & 
\multicolumn{2}{c|}{$[0.14,  0.67]$} \\
    & DW/BOX & $[0.55,  0.72]$ & $[0.58,  0.83]$ & 
\multicolumn{2}{c|}{$[0.14,  0.76]$} \\\hline
    & TH/FULL &   $\geq 0.86$  &     --           &
\multicolumn{2}{c|}{       -- }       \\
Strong & TH/BOX  & $\geq 0.66$  & $[0.53,  0.57]$\;$\oplus$ \;$\geq 0.68$ &    
\multicolumn{2}{c|}{--}           \\
    & DW/FULL &  $\geq 0.86$    &       --         &      
\multicolumn{2}{c|}{ --}        \\
    & DW/BOX &  $\geq 0.72$    &  $\geq 0.83$  &       
\multicolumn{2}{c|}{ -- }      \\\hline
\end{tabular}}
\caption{Ranges of
$\dNeff$ for which we find the evidence against or in favour of the 3+1 model 
compared to the model with only the 3 active neutrinos 
to  be weak, moderate or  strong. }
\label{tab:favranges}
\end{table} 

The second test performed deals with the (in)compatibility of the sterile mass
constraints as required to describe SBL and cosmology.  For this we
have evaluated the $\Delta\chi^2$ defined in \eref{eq:chi2comp} 
which we plot in the right panels in \figref{fig:Bupf-chi2}.   
Altogether we read that this test yields 
inconsistency on the  $m_s$ required by cosmology and SBL larger than 
$3\sigma$ for 
\begin{eqnarray}
&&\dNeff\geq 0.45\, (0.54)\, [0.76]\, \left([0.82]\right) \;\; {\rm For\, CMB+BAO}  \:, \nonumber \\
&&\dNeff\geq 0.39\, (0.50)\, [0.65]\, \left([0.77]\right) \;\; {\rm For\, CMB+BAO+BICEP2} \;,   \\
&&\dNeff\geq 0.56\, (0.67)\, [0.83]\, \left([0.89]\right) \;\; {\rm For\, CMB+BAO+BICEP2+HST+PlaSZ} \;,   \nonumber
\end{eqnarray}
for thermal (WP) $\nu_s$ scenario for the full [box] SBL likelihood.  

In summary, we find that the analysis of  cosmological results from temperature 
and polarization data on the CMB as well as from the BAO measurements 
from LSS data disfavours the 3+1 sterile models introduced to explain the
SBL anomalies over the scenario without sterile neutrinos, and also that 
their allowed/required ranges of $m_s$ are incompatible. This is so
even if new physics is involved so that the contribution of the sterile 
neutrino to the energy density of the Universe 
(and therefore to the cosmological observables) is suppressed 
with respect to that of the fully thermalized case resulting from its
mixing with the active neutrinos. 
When the local measurement
of the $H_0$ by the Hubble Space Telescope,  and the cluster SZ cluster data 
from the Planck mission is included, compatibility can be found between 
cosmological and SBL data, but still requires a substantial suppression
of the $\nu_s$ contribution to  $\rho_R$.

%%%%%%%%%%%%%%%%%%%%%%%%%%%%%%%%%%%%%%%%%%%%%%%%%%%%%%%%%%%%%%%%%%%%%%

\section*{Acknowledgments}
This work is  supported by USA-NSF grants PHY-09-69739 and PHY-1316617, by
CUR Generalitat de Catalunya grant 2009SGR502 by MICINN FPA2010-20807 
and consolider-ingenio 2010 program grants CUP (CSD-2008-00037) and CPAN,
and by EU grant 
FP7 ITN INVISIBLES (Marie Curie Actions PITN-GA-2011-289442).
J.S. acknowledges support from the Wisconsin IceCube Particle
Astrophysics Center (WIPAC) and U. S. Department of Energy under the
contract DE-FG-02-95ER40896.
%%%%%%%%%%%%%%%%%%%%%%%%%%%%%%%%%%%%%%%%%%%%%%%%%%%%%%%%%%%%%%%%%%%%%%

\clearpage

%%%%%%%%%%%%%%%%%%%%%%%%%  Bibliography    
%%%%%%%%%%%%%%%%%%%%%%%%%%%%%%%%%%%%%%%%%%%%%%%%%%%%%%%%%%%%
\bibliography{biblio}{}
\bibliographystyle{JHEP}

%%%%%%%%%%%%%%%%%%%%%%%%%%%%%%%%%%%%%%%%%%%%%%%%%%%%%%%%%%%%%%%%%%%%%%%%%%%%%%%%%%%%%%%%%%%%%%%%%%%%%%%%%%
%%%%%%%%%%%%%%%%%%%%%%%%%%%%%%%%%%%%%%%%%%%%%%%%%%%%%%%%%%%%

\end{document}